# THE RADIATIVE NEUTRON CAPTURE ON $^2$H, $^6$Li, $^7$Li, $^{12}$C AND $^{13}$C AT ASTROPHYSICAL ENERGIES

SERGEY DUBOVICHENKO[*,†], ALBERT DZHAZAIROV-KAKHRAMANOV[*,†] and NATALIA BURKOVA[‡]

[*]*V. G. Fessenkov Astrophysical Institute "NCSRT" NSA RK, 050020, Observatory 23, Kamenskoe plato, Almaty, Kazakhstan*
[†]*Institute of nuclear physics NNC RK, 050032, Almaty, Kazakhstan*
[‡]*Al-Farabi Kazakh National University, 050040, Almaty, Kazakhstan*
[*]*dubovichenko@mail.ru*
[†]*albert-j@yandex.ru*
[‡]*natali.burkova@gmail.com*

The continued interest to the study of the radiative neutron capture on atomic nuclei is caused, on the one hand, by the important role of this process in the analysis of many fundamental properties of nuclei and nuclear reactions, and, on the other hand, by the wide use of the capture cross section data in the various applications of nuclear physics and nuclear astrophysics, and, also, by the analysis of the processes of primordial nucleosynthesis in the Universe. This review is devoted to description of the results obtained for the processes of the radiative neutron capture at thermal and astrophysical energies on certain light atomic nuclei. The consideration of these processes is done in the frame of the potential cluster model, the general principles of which and calculation methods were described earlier. The methods of usage of the obtained on the basis of the phase shift analysis intercluster potentials will be directly demonstrated for calculations of the radiative capture characteristics. The considered capture reactions are not a part of stellar thermonuclear cycles, but they get in the basic reaction chain of primordial nucleosynthesis, taken place in the time of the Universe formation.

## 1. Introduction

Earlier, we have shown the possibility to describe the astrophysical *S*-factors[1,2] of the radiative capture reactions on numerous light and lightest atomic nuclei in the frame of the potential cluster model (PCM) with the forbidden states (FS).[3-5] This model takes into account the supermultiplet symmetry of the cluster system wave function with the separation of orbital states according to Young schemes.[3,4,6] The using classification of the orbital states allows to analyze the structure of intercluster interactions, to determine the existence and the number of allowed states (AS) and forbidden states in the intercluster potentials, and, consequently, gives the possibility to find the number of nodes of the radial wave function (WF) of the relative cluster motion.[7-9] In this approach, the potentials of intercluster interactions for scattering processes are constructed on the basis of description of elastic scattering phase shifts derived from the experimental differential cross sections during the phase shift analysis.[3,4,10]

The potentials for the bound states (BS) of light nuclei in cluster channels are constructed not only on the basis of the description of scattering phase shifts, as some additional requirements are used. For example, the reproducing of the binding energy and some other characteristics of the ground states (GS) of nuclei is such a demand, at that in some cases this requirement is the main.[3-5] At the same time, it is assumed that the BS is caused, in general, by the cluster channel consisted of initial particles, taking part in the reaction.[11-13] In addition, for any nucleon system, the many-particle character of the problem is taken into account by the separation of single-particle levels of such a potential to the allowed and forbidden by the Pauli principle states.[14] The selection of the potential cluster model for the consideration of such cluster systems in nuclei, nuclear and thermonuclear processes at astrophysical energies[2,15] is caused by the fact that the probability of formation of nucleon associations, i.e., clusters and degree of their isolation from each other are relatively high in many light nuclei. This is confirmed by the numerous experimental measurements and different theoretical calculations obtained by various authors in the last fifty-sixty years.[3,14,16,17]

In the beginning of this review we will consider the possibility to describe the total cross sections of the neutron capture on $^2$H on the basis of the potential cluster model, where the supermultiplet symmetry of wave function and the separation of orbital states according to Young schemes are taken into account. Although, the n$^2$H → $^3$Hγ radiative capture reaction at astrophysical energies with the formation of unstable tritium nucleus, which is turned to $^3$He due to β-decay, is not a part of basic thermonuclear cycles,[2] it evidently can play a certain role in some models of Big Bang.[18-23] It is assumed, that in these models the primordial nucleosynthesis goes, for example, according to the basic nuclear reaction chain of the form:

$$^1H(n,\gamma)^2H(n,\gamma)^3H(^2H,n)^4He(^3H,\gamma)^7Li(n,\gamma)^8Li(^4He,n)^{11}B(n,\gamma)^{12}B(\beta^-)^{12}C(n,\gamma)$$
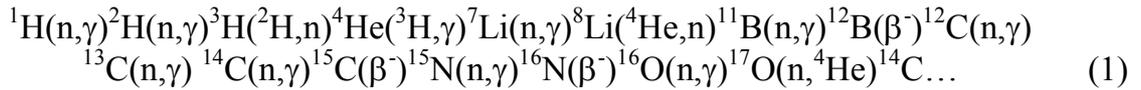
$$^{13}C(n,\gamma)\ ^{14}C(n,\gamma)^{15}C(\beta^-)^{15}N(n,\gamma)^{16}N(\beta^-)^{16}O(n,\gamma)^{17}O(n,^4He)^{14}C\ldots \qquad (1)$$

etc.[18-23] The neutron capture on $^2$H is a part of this chain and not only at low energies, those will be considered below. Note that there are other variants of records of the chain[18-23] with other transitional reactions. In addition, the considered n$^2$H → $^3$Hγ capture reaction is the mirror one to the p$^2$H → $^3$Heγ process, which was considered in our previous works.[3,4,24-26] The last process is a part of thermonuclear proton-proton chain, be its first reaction that goes due to electromagnetic interactions. Evidently, this cycle provides the main energy release of the nuclear reactions[27] that are the cause of burning of the Sun and stars of our Universe.

Passing to the consideration of the capture reaction in the n$^6$Li system, notes that the phase shift analysis of the p$^6$Li elastic scattering was done in our previous work Dubovichenko *et al.*[28] at the energy range 500÷1150 keV,[29,30] that is interesting for nuclear astrophysics. The potentials for the ground $^2P_{3/2}$ and for the first excited $^2P_{1/2}$ state of $^7$Be in the p$^6$Li channel, as well as the potential of the doublet $^2S_{1/2}$ scattering wave were constructed. The obtained results allow to consider the astrophysical S-factor of the radiative proton capture on $^6$Li at low energies.[4,31] As the result, it was shown that the used approach allows to describe the existent experimental data for the processes of the radiative proton capture on $^6$Li in a wide energy range.[32-34]

Further, developing the obtained results, we will stop on the consideration of total cross sections of the radiative neutron capture on $^6$Li. Although this reaction, evidently, is of certain interest for nuclear astrophysics,[18-23] it is inefficiently studied experimentally. This conclusion is made up from the consideration of the experimental results from databases of MSU[35] or EXFOR.[36] According to Refs. 35, 36 there are only two types of measurements, carried out at 0.025 eV (25 meV)[37-40] and in the range 6.7÷7.3 MeV, where there are overlapped resonances at 13.7 and 17 MeV, relative to the ground state of $^7$Li, with the width about 0.5 MeV, but with more undetermined yet characteristics, for example momentum and parity.[41] This, by-turn, does not allow to include them into the analysis of possible electromagnetic transitions to the ground and first excited states of $^7$Li, which are formed as a result of the treated here capture reaction n$^6$Li → $^7$Liγ. Nevertheless, it is interesting to consider the possibility of description of the total cross sections of this reaction at thermal and astrophysical energy range from 0.025 eV to 1÷2 MeV. We have used here, as earlier in the case of p$^6$Li system,[3] the potential cluster model with forbidden states and classification of cluster states according to Young schemes.[42]

Then, radiative capture reaction n$^7$Li → $^8$Liγ will be considered at astrophysical energies with the formation of β-active $^8$Li. This reaction also directly does not take part in the basic thermonuclear cycles,[2] but it can play the essential role in certain models of the Big Bang (see Eq. (1)).[18-23] In addition, the considered reaction is the mirror reaction relative to the p$^7$Be → $^8$Bγ capture, where $^8$B decays to $^8$Be+e$^+$+ν because of the weak process. Neutrinos in this reaction have the relatively big energy and are registered in earth conditions already over a period of few decades, and unstable $^8$Be nucleus decays into two α particles. The p$^7$Be → $^8$Bγ capture reaction is one of the final processes of thermonuclear proton-proton chain, which, as it usually assumed, causes the burning of the Sun and majority of stars of our Universe.[2-4]

Finally, we will consider the reactions of neutron capture on $^{12}$C and $^{13}$C at thermal and astrophysical energies, which are part of the basic chain of thermonuclear reactions of the primordial nucleosynthesis of Eq. (1).[18-23] The available experimental data, for example, on total cross sections of the n$^{12}$C reaction are given in works Refs. 43-50 and can be found in data bases.[35,36] These data give the general representation about the form of radiative capture cross sections in a wide energy range, though they do not cover the whole energy region. Therefore, it interesting to clarify the possibility of description of these cross sections on the basis of the PCM with FS, as it was done earlier for the radiative proton capture on $^{12}$C and $^{13}$C.[4,5,51,52] Note, that our newly done phase shift analysis, including new experimental data on differential cross sections of the p$^{12}$C and p$^{13}$C elastic scattering at astrophysical energies,[2,53,54] has made it possible to construct the quite unambiguous potentials of p$^{12}$C and p$^{13}$C interactions according to the obtained elastic scattering phase shifts. They, in general, should not considerably differ from the analogous potentials of the n$^{12}$C and n$^{13}$C scattering, and bound states of $^{13}$C in n$^{12}$C and $^{14}$C in n$^{13}$C channels.

## 2. Radiative neutron capture on $^2$H in cluster model

Possibility to describe the experimental data on total cross sections of the radiative neutron capture on $^2$H at thermal (~1 eV), astrophysical (~1 keV), and low (~1 MeV) energies will be considered in the frame of the potential cluster model with forbidden states and their classification according to Young schemes. It was shown that the used model and the developing here numerical methods of its realization are able to describe correctly the behavior of the experimental cross sections at the energy range from 10 meV (10·10$^{-3}$ eV) to 15 MeV.

## 2.1. *Potential description of the n²H elastic scattering*

Before consideration of the n²H system, let us briefly stay on the results obtained earlier for the p²H scattering process.[3-5] The potentials of the p²H elastic scattering for each partial wave are constructed, so that to correctly describe corresponding partial phase shifts of elastic scattering at low energies,[55-58] which are mixed according to Young schemes {3} + {21} in the doublet channel.[3,4,6] Using this conception, we have obtained the p²H potentials for scattering processes, which are mixed according to Young schemes {3} and {21} and are represented as

$$V(r) = V_0 \exp(-\gamma r^2) + V_1 \exp(-\delta r), \qquad (2)$$

with parameters listed in Table 1.[59,60]

Then, the pure phase shifts with the scheme {3} were separated in the doublet spin channel and the pure according to Young schemes potentials of the $^2S$ intercluster interaction of the ground state of ³He in the p²H channel are constructed on their basis, the parameters of these potentials are given in Table 1 or, for example, in our works from Refs. 3, 4, 6, 59-61. The parameters of this potential leads to the relatively good description of the main characteristics of ³He in the p²H channel (see, for example, Refs. 3-6, 10, 24-26, 60, 61).

Table 1. The doublet potentials of the p²H interaction.[16] $E_{BS}$ is the energy of the ground bound state of ³He in the p²H channel, $E_{exp}$ – its experimental value.

| $^{(2S+1)}L, \{f\}$ | $V_0$ (MeV) | $\gamma$ (fm⁻²) | $V_1$ (MeV) | $\delta$ (fm⁻¹) | $E_{BS}$ (MeV) | $E_{exp}$ (MeV) |
|---|---|---|---|---|---|---|
| $^2S, \{3\}+\{21\}$ | -55.0 | 0.2 | – | – | – | – |
| $^2P, \{3\}+\{21\}$ | -10.0 | 0.16 | +0.6 | 0.1 | – | – |
| $^2S, \{3\}$ | -41.55562462 | 0.2 | – | – | -5.493423 | -5.493423 |

The calculations of the total cross sections of the radiative proton capture on ²H and the astrophysical *S*-factors at energies down to 10 keV were done with these potentials,[60] although at that moment we knew only the experimental data on *S*-factor at the energy range above 150÷200 keV.[62] Later, the new experimental results at energies down to 2.5 keV are appeared.[63-65] After their consideration, it was found that the previous calculations, which based on the *E*1 process only, completely agree with them in the range from 1 MeV down to 10 keV.[60] Thereby, the using potential cluster model allows not only to describe new data, but, intrinsically, to predict the behavior of the astrophysical *S*-factor of the proton capture on ²H in the energy range up to 10 keV. As a result, the calculations, presented in 1995 in our work,[60] were done before the carrying out of the new experimental measurements[65] in 2002 and even before the earliest works,[63,64] published in 1997.

Here, we will use the obtained in Refs. 3, 6, 16, 59, 60, 66 p²H potentials for the consideration of the radiative neutron capture on ²H at low energies, using, at once, the same calculation methods, which were checked for the p²H system.[3] The parameters of the GS potential of ³H in n²H channel without Coulomb interaction were slightly improved for correct description of the bound energy of tritium, which is equal to -6.257233 MeV.[67,68] As a result, the following parameters of the potential

of Eq. (2) were obtained

$$V_0 = -41.4261655 \text{ MeV and } \gamma = 0.2 \text{ fm}^{-2}. \quad (3)$$

This potential reproduces the binding energy of $^3$H accurately, giving the value -6.257233 MeV, it yields the charge and mass radii 2.33 and 2.24 fm, respectively, using the charge neutron radius equals zero, its mass radius equals proton radius 0.8775(51) fm and at the deuteron radius 2.1424(21) fm.[69] The asymptotic constant (AC), defined as in Ref. 70

$$\chi_L(r) = \sqrt{2k_0} C_w W_{-\eta L+1/2}(2k_0 r), \quad (4)$$

is equal to 2.04(1) at the interval 5÷15 fm. The AC error is formed by its averaging over the mentioned interval, but its values, obtained in different works, are given in Ref. 70 and are in the range 1.82÷2.21. In this expression $\chi_L(R)$ is the numerical wave function of the bound state obtained from the solution of the radial Schrödinger equation and normalized to unity; $W_{-\eta L+1/2}$ is the Whittaker function of the bound state determining the asymptotic behavior of the WF. It is the solution of the same equation without nuclear potential, i.e., the long distance solution; $k_0$ is the wave number determined by the channel binding energy; $\eta$ is the Coulomb parameter; $L$ is the orbital moment of the bound state.

Notes, that the given above value of binding energy was obtained at the calculation accuracy of finite-difference method (FDM) of $10^{-6}$ MeV, and, using the multiple accuracy of 2 $10^{-9}$, it is possible to obtain more accurate value -6.257233014 MeV. In addition, since deuteron has the radius more than tritium 1.755(86) fm,[67] it can not be inside tritium in free, i.e., not deformed state, and the degree of its deformation, as it was shown in Ref. 71, is equal near 30%.[16] The same conclusion is in Ref. 72, where it was shown that the WF of deuteron located in tritium drops faster than the WF of deuteron in its free state. Thereby, the existence of the third particle, neutron in this case, leads to the deformation, i.e., compression of the deuteron cluster inside tritium nucleus. Approximately the same conclusion was done in the calculations using resonating group method (RGM); the analysis of these results was done in Ref. 73 and the usual estimation of the deuteron deformation is about 20÷40%.

Two-particle variational method (VM) with the expansion of relative cluster motion WF by non-orthogonal Gaussian basis and the independent variation of parameters[3,61] is used for additional check of the obtaining of the binding energy of $^3$H in such potential, i.e., for the n$^2$H bound state with the interaction of Eq. (3)

$$\Phi_L(R) = \frac{\chi_L(R)}{R} = R^L \sum_i C_i \exp(-\alpha_i R^2), \quad (5)$$

where $\alpha_i$ and $C_i$ are the variational parameters and expansion coefficients.

The variational method allows to obtain the binding energy of -6.2572329999 MeV ≈ -6.257233000 MeV by using independent variation of parameters and the Gaussian basis having dimension $N = 10$. The asymptotic constant $C_W$ of the variational WF with parameters given in Table 2, remains at the level of 2.05(2) at distances of 6÷20 fm that is not differ from the FDM value, and the residual errors are not more than $10^{-11}$.[61]

Table 2. The variational parameters and expansion coefficients of the bound state WF of $^3$H in the n$^2$H system.

| $i$ | $\alpha_i$ | $C_i$ |
| --- | --- | --- |
| 1 | 3.361218182141637E-001 | 1.231649877959069E-001 |
| 2 | 2.424705040532388E-002 | 1.492826524302106E-002 |
| 3 | 1.168704181683766E-002 | 1.190880013572610E-003 |
| 4 | 9.544908567362362E-002 | 1.304076551702031E-001 |
| 5 | 867951954385213E-002 | 5.868193953570694E-002 |
| 6 | 9.341901487408062E-001 | -2.155090483420204E-002 |
| 7 | 1.756025156195464E-001 | 1.814952898311890E-001 |
| 8 | 2.396705577261060E-001 | 6.944804259139825E-002 |
| 9 | 6.503621155681423E-001 | 1.564362603986158E-002 |
| 10 | 9.684977093058702E-001 | 1.709621746273126E-002 |

*Note.* Normalization coefficient of the wave function on the interval of 0÷25 fm is $N$ = 9.999999996433182E-001.

It is known that the variational energy decreases as the dimension of the basis increases and gives the upper limit of the true binding energy. At the same time the finite-difference energy increases as the size of steps decreases and the number of steps increases.[3-5,61] Therefore, for the real binding energy in this potential it is possible to use the average value, obtained above on the basis of two used methods, and equals -6.257233007(7) MeV for the n$^2$H system. Thereby, we obtain that the accuracy of determination of the binding energy of this system in the listed above BS potential of Eq. (3) and obtained by two different methods (VM and FDM), on the basis of two different computer programs[5,61] is on the level of ±0.007 eV or ±7 meV.

## 2.2. *The total cross sections of the radiative neutron capture on $^2$H*

At first, we will show the working capacity of the potential cluster model used here, potentials obtained on the basis of the p$^2$H elastic scattering phase shifts, and the procedure of separation of the pure phase shift and corresponding the GS potential of $^3$H on example of the photodisintegration of $^3$H into n$^2$H channel. It was considered by us earlier in Ref. 60 in more wide energy region, but less thoroughly. The results of these calculations at the energies of γ-quanta 6.3÷10.5 MeV are shown in Fig. 1 by the solid line for the sum of the $E$1 and $M$1 cross sections with the given above p$^2$H potentials (see Table 1) with switch off Coulomb interaction.

The contribution of $M$1 process to disintegration of $^3$H into the $^2S$ doublet wave of n$^2$H scattering, which does not give the appreciable contribution to the cross sections at these energies, is shown in Fig. 1 by the dashed line. The cross sections of the considered process are caused exclusively by the $E$1 transition at the decay of the GS of $^3$H into the doublet $^2P$ scattering wave. The experimental data for total cross sections of the photodisintegration reaction of $^3$H into the n$^2$H channel for considered energies were taken from works: Ref. 74 – black points,

Ref. 75 – black triangles.

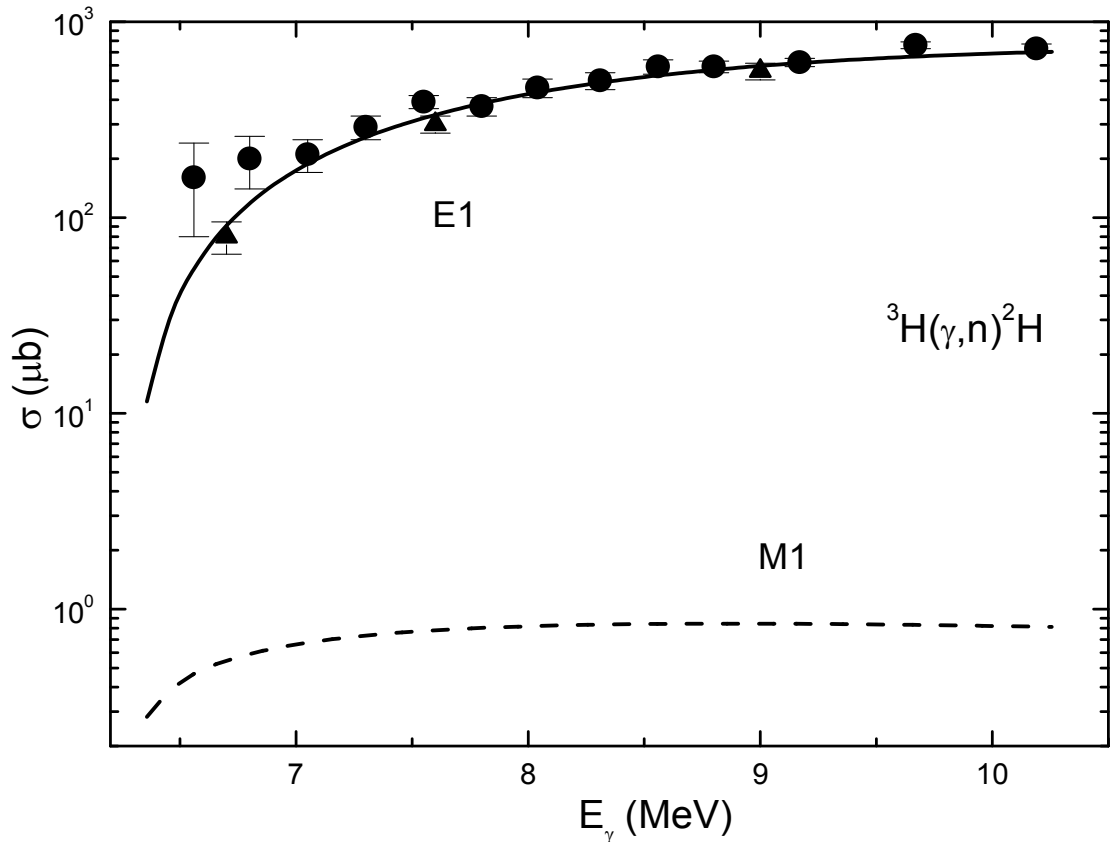

Fig. 1. The total photodisintegration cross sections of $^3$H into the n$^2$H channel. The experimental data are from: ▲ – Ref. 75 and • – Ref. 74. Explanations for lines are in the text.

Further, the cross section calculations of the radiative neutron capture on $^2$H at the energy range 10 meV÷15 MeV were done with the same parameters of the p$^2$H nuclear potentials for the $^2S$ and $^2P$ scattering waves and for the GS from Eq. (3) without Coulomb component. The results, which are shown in Fig. 2, demonstrate the prevalence of the $M$1 process at the energies lower than 1 keV, which cross section is shown by the dashed line. The dotted line shows the contribution of the $E$1 transition, and the solid line represents the summarized cross sections of the $E$1 and $M$1 transitions. The experimental data for total cross sections of the radiative neutron capture on $^2$H are taken from the works: Ref. 76 – points at energies 30, 55 and 530 keV, Ref. 77 – circles at 7÷14 MeV, Ref. 78 – triangle at 0.01 eV, Ref. 44 – asterisk at 0.025 eV, Ref. 79 – square at 50 keV.

As it is seen from Fig. 2, the cross section of the $E$1 transition represented by the dotted line drops sharply and already at 0.1 keV it can be neglected. At the same time, this process at the region above 10 keV is dominant and absolutely determines the behavior of total cross sections, which allow us to describe the existent experimental data at energies from 50÷100 keV to 15 MeV. At lower energies, approximately from 30 keV to 0.01 eV (10 meV), the calculated cross sections have the value slightly less than measured in the experiments.[44,76,78] The calculation of the cross section at 0.01 eV gives the value of the cross section that is, approximately, 1.5 times less than the experiment.

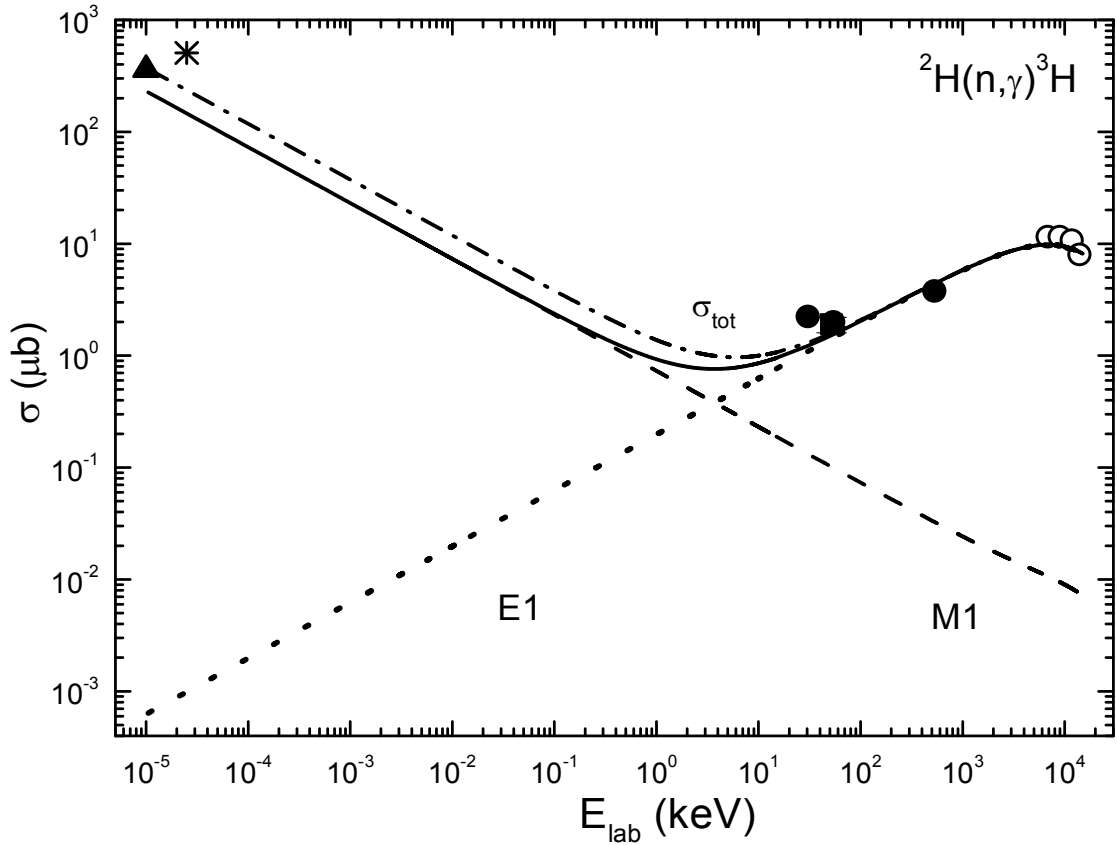

Fig. 2. The total cross section of the neutron radiative capture on $^2$H. The experiment from: Ref. 76 – •, Ref. 77 – ○, Ref. 78 – ▲, Ref. 44 – *, Ref. 79 – ■. Explanations for lines are in the text.

However, the obtained earlier p$^2$H potential without Coulomb interaction for the $^2S$ scattering wave was used here for calculation of the $M$1 transition. The data spread for different phase shifts obtained from the experimental data for the p$^2$H elastic scattering[55-58] reaches 10÷20%, what is shown in Fig. 3 by points. Therefore, even the p$^2$H scattering potential, which phase shift is shown in Fig. 3 by the dashed line, is constructed on their basis with big ambiguities, but here, we are considering the n$^2$H system, for what we did not succeeded in finding the results of phase shift analysis in the considered energy range.

Therefore, further we will consider the required changes, which will be necessary for the p$^2$H potential in the $^2S$ scattering wave, so that the result will be possible to describe the existent experimental data at lowest energies. Let us note that these changes wouldn't affect for the results of the $E$1 process, which plays the main role at relatively high energies, as it was shown in Fig. 2.

Consequently, the results for the total cross section of the radiative capture shown in Fig. 2 by the dashed-dot line were obtained. The depth of the $^2S$ potential in the n$^2$H elastic scattering is not much higher than for the p$^2$H system from Table 1

$$V_0 = -60.0 \text{ MeV and } \gamma = 0.2 \text{ fm}^{-2}. \qquad (6)$$

The scattering phase shift, obtained for this potential, is shown in Fig. 3 by the solid line. Evidently, the $^2S$ phase shift of the improved n$^2$H potential at low, up to 3 MeV, energies drops quicker than the analogous phase shift for the p$^2$H potential that was obtained without taking into account Coulomb interaction. This fact, by-turn, affects the calculation results of the total cross sections for the *M*1 process and, as it is seen from Fig. 2, the usage of this potential allows to get a good description of the existent data for total cross sections, even at the lowest energies.

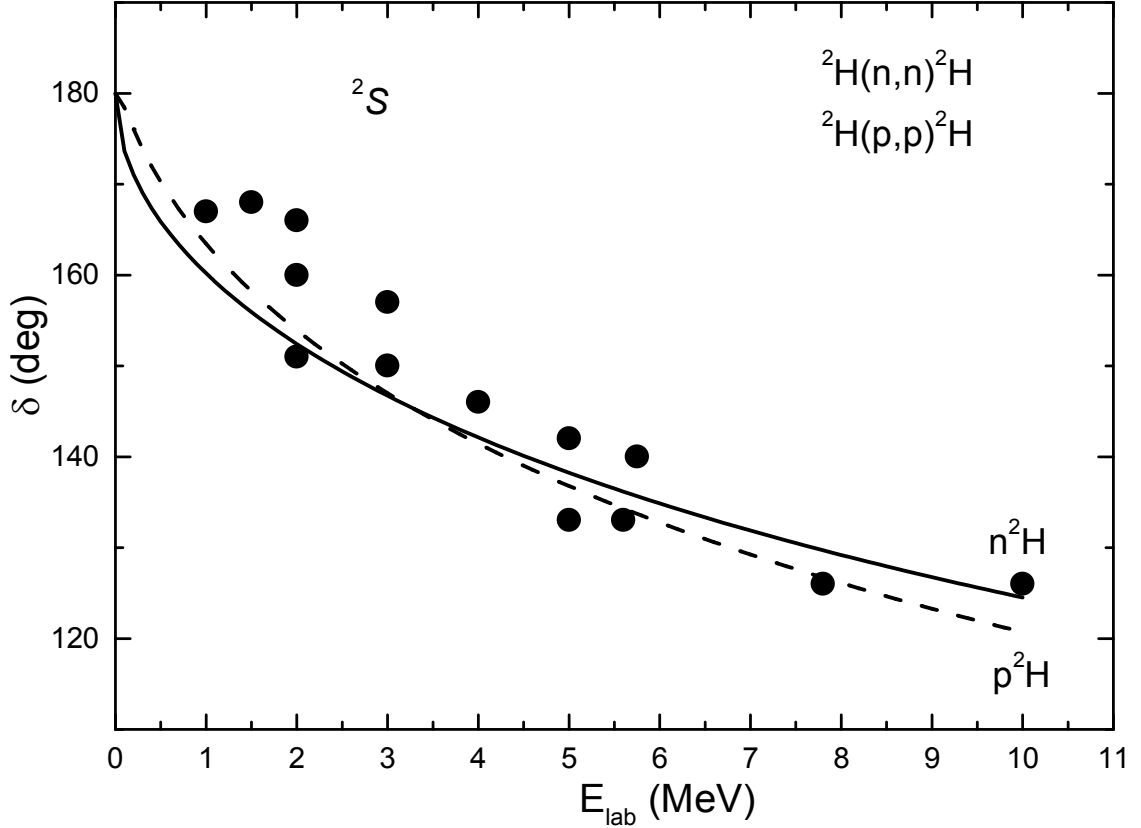

Fig. 3. The $^2S$ phase shifts of the p$^2$H (dashed line) and n$^2$H (solid line) elastic scattering. Points: • – the phase shifts obtained from the experimental data in Refs. 55-58. The potential parameters are given in the text and in Table 1.

Thereby, the change of parameters of the n$^2$H potential in the $^2S$ phase shift less than 10% immediately allows to describe the existent experimental data at low energies. Such change of the parameters can be interpreted by the ambiguity of the existent p$^2$H phase shifts and their absence for the n$^2$H elastic scattering. Consequently, the using potential cluster model has allowed to reproduce correctly the experimental data for the total cross sections of the radiative neutron capture on $^2$H at the energy range, when energies at the edges of diapason differ from each other by more than nine orders, notably from $10^{-5}$ keV to $1.5 \cdot 10^4$ keV.

Since, the calculated cross section, which is shown in Fig. 2 by the dashed-dot line, is practically the straight line at energies from $10^{-5}$ to 0.1 keV, so it can be approximated by the simple function of the form

$$\sigma_{ap}(\mu b) = \frac{1.1830}{\sqrt{E_n(\text{keV})}} \ . \qquad (7)$$

The value of given constant 1.1830 μb keV$^{1/2}$ was determined from the one point of the cross sections at the minimal energy, equals $10^{-5}$ keV. Further, it is possible to consider the absolute value of the relative deviation of the calculated theoretical cross section and the approximation of this cross section by this function in the range from

$10^{-5}$ to 0.1 keV

$$M(E) = \left|[\sigma_{ap}(E) - \sigma_{theor}(E)]/\sigma_{theor}(E)\right|. \qquad (8)$$

It was found that at the energy range lower 100 eV this deviation does not exceed 1.5÷2.0%. It is possible, evidently, to suppose that the shape of the dependence of total cross section from energy from Eq. (7) will be also preserve at lower energies. In this case, the estimation of the value of total cross section, for example at the energy 1 μeV ($10^{-6}$ eV = $10^{-9}$ keV), gives the value 37.4 mb.

## 3. The radiative neutron capture on $^6$Li

Let us consider the possibility to describe the experimental data for the total cross sections of the radiative neutron capture on $^6$Li at the energy range from 25 meV to 1.0÷3.0 MeV in the frame of the cluster model with forbidden states and classification of cluster states according to Young schemes. At once notes, that it is necessary to take into account only the $E1$ transition, from the $^2S_{1/2}$ state of the n$^6$Li scattering to the $^2P_{3/2}$ and first excited $^2P_{1/2}$ states of $^7$Li in the final channel, for the acceptable description of the existent data.

### 3.1. *Potential description of the n$^6$Li elastic scattering*

Even if this reaction, evidently, is of certain interest for some problems of nuclear astrophysics[1,18-23] with a view to formation and accumulation of the lithium isotopes, it was experimentally studied relatively few. According to the data[35,36] there are only measurements carried out at 0.025 eV (25 meV),[37-40] and also for three values for energy range 30÷80 keV from.[80] Besides, in data bases[35,36] and works[81-83] there are data for the total cross sections of photodisintegration of the GS of $^7$Li into n$^6$Li channel (see Fig. 4), which were recalculated here to the capture cross sections in the range 0.05÷1.5 MeV. As far as the disintegration is going from the GS of $^7$Li, the principle of detailed balancing with $J_0$=3/2 and $J_0$=1/2 at the identical disintegration cross sections is used for estimation of the value of summarized capture cross section to the ground and first excited states.

$$\sigma_c(3/2+1/2) = \sigma_c(3/2)+\sigma_c(1/2) = 4A(q,K)\sigma_d(3/2)+2A(q,K)\sigma_d(3/2), \qquad (9)$$

where

$$\sigma_c(J_0) = (2J_0+1)\frac{2K^2}{q^2(2S_1+1)(2S_2+1)}\sigma_d(J_0) = (2J_0+1)A(q,K)\sigma_d(J_0) \qquad (10)$$

and $J_0$ – total moment of the bound state of the nucleus, $\sigma_c$ – total cross sections of the radiative capture, $\sigma_d$ – total cross sections of the direct photodisintegration. The results of such recalculation are shown in Fig. 5 by the circles, open and black squares. Since all these data well determine the general behavior of the total capture cross sections, it will be interesting to consider the possibility of their theoretical description in the energy range from 0.025 eV to 1÷1.5 MeV using for this, as before in the case of p$^6$Li system,[3,4] the potential cluster model with FS and carried out classification of the cluster states according to orbital Young schemes.[42]

Further, it was noted in Refs. 3, 5, 6, 42 that, generally speaking, it can be two variants of potentials for the $^2S$ and $^2P$ waves in the $N^6$Li system. In the first case, there are two BS in these partial waves and only one of them in the $^2P$ waves is allowed and corresponding to the GS of the nuclei with A = 7, and all other states are forbidden. In the second case, these waves contain one BS – in the $^2S$ wave it is forbidden, and in the $^2P$ wave it corresponds to the allowed BS $^2P_{3/2}$ and $^2P_{1/2}$. Therefore, further we will consider both variants of potentials for the $^2S$ scattering states and $^2P$ bound states of $^7$Li in the n$^6$Li channel. At that, only the variants of the potentials that can lead to the acceptable description of the total cross sections of the radiative neutron capture on $^6$Li, elastic scattering phase shifts and main characteristics of the BS.

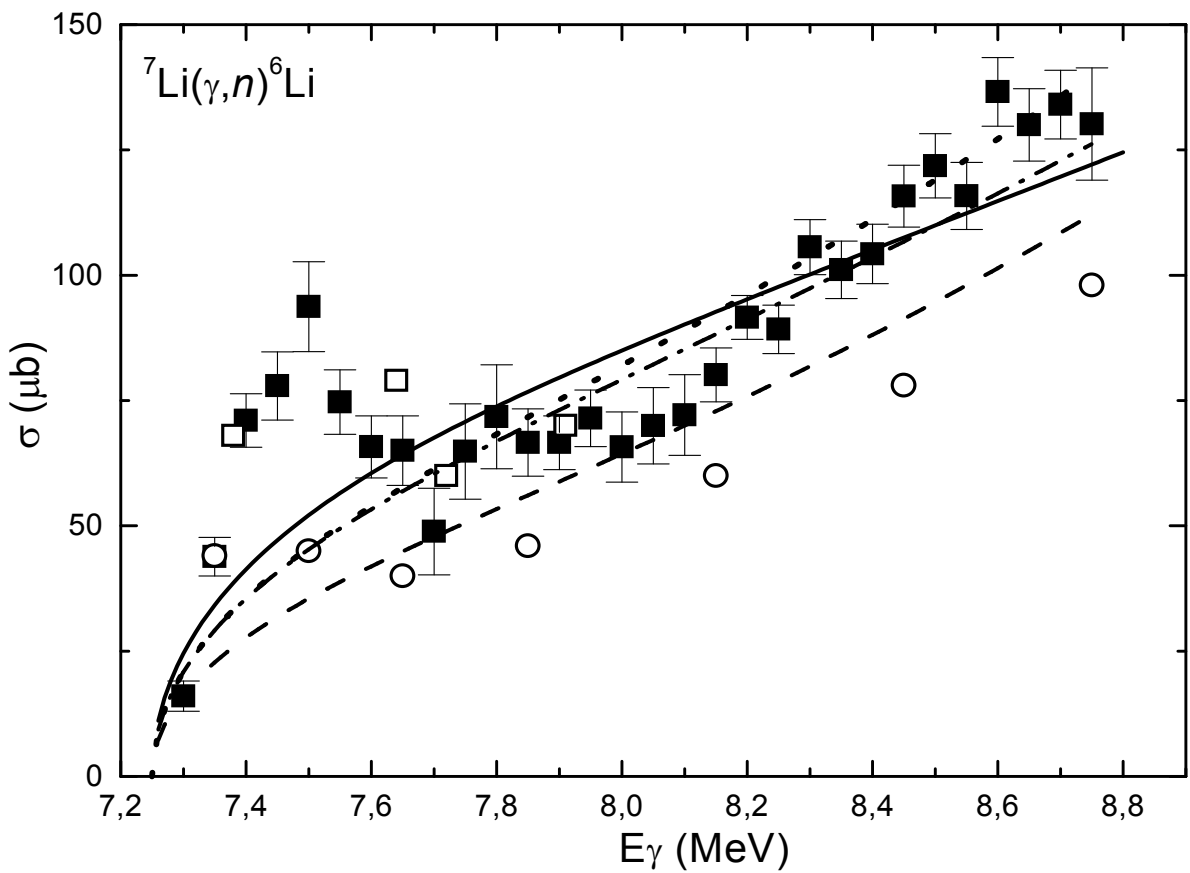

Fig. 4. The total cross sections of the reaction $^7$Li($\gamma$, n)$^6$Li at low energies. The experimental results of works: ■ – Ref. 81, ○ – Ref. 82 and □ – Ref. 83. Lines – the calculation with the potentials listed in the text.

Starting the consideration of the n$^6$Li → $^7$Li$\gamma$ radiative capture reaction, note that originally the phase shift analysis of the p$^6$Li scattering with taking into account the spin-orbital splitting was done at the energy range from 0.5 to 5.6 MeV in Ref. 84. Later, these results for the $S$ scattering phase shifts were slightly improved in Ref. 28 and were used for construction of the intercluster potentials, which are used in the calculations of the astrophysical $S$-factor of the radiative proton capture on $^6$Li.[31]

At first, we will use obtained earlier p$^6$Li interaction potentials, but already without Coulomb term[3,6,28] and will consider the total cross sections of the radiative neutron capture on $^6$Li at the range of astrophysical energies on the basis of the PCM. The doublet $^2S_{1/2}$ Gaussian potential ($V_1 = 0$) with the parameters

$$V_S = -124 \text{ MeV and } \gamma_S = 0.15 \text{ fm}^{-2}, \qquad (11)$$

is preferable for the description of our results for the scattering phase shifts of the p$^6$Li elastic scattering, as it was shown in Refs. 28, 31, and this potential contains two forbidden bound states, corresponding to orbital Young schemes {52} and {7}.[3,6,31] The phase shifts of such potential for the p$^6$Li elastic scattering are shown in Fig. 6 by the dashed line, and for the n$^6$Li scattering, i.e., without Coulomb term – by the solid line.

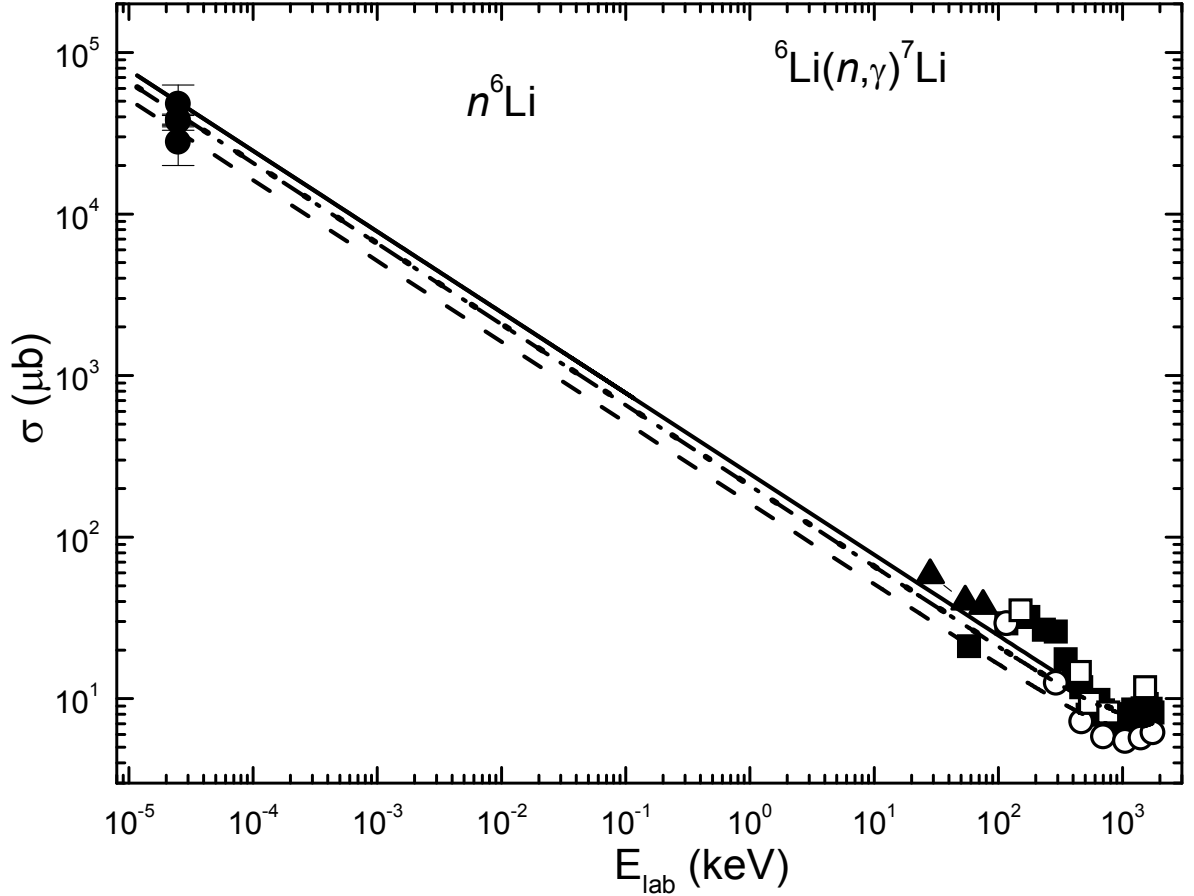

Fig. 5. The total cross sections of the radiative neutron capture on $^6$Li at low energies. Experimental results: ● – Refs. 37-40, ▲ – Ref. 80, ■ – Ref. 81, ○ – Ref. 82, □ – Ref. 83. Lines – the calculation of the total cross sections with the potentials listed in the text.

The phase shifts for other variant of the shallow $^2S$ potential, which contains only one FS and has the parameters

$$V_0 = -34 \text{ MeV and } \gamma = 0.15 \text{ fm}^{-2}, \text{ at } V_1=0 \qquad (12)$$

are shown in Fig. 6 by the dotted line for the p$^6$Li scattering and by the dashed-dot line for the n$^6$Li scattering. Our results obtained in Ref. 28 are shown in Fig. 6 by the points in the capacity of the p$^6$Li scattering phase shifts that are extracted from the experiment. As it is seen from Fig. 6, both of these potentials leads to the identical description of the p$^6$Li phase shifts, and the n$^6$Li phase shifts have a few differ at low energies.

The $^2P_{3/2}$ wave potential of the ground state of $^7$Be [31], which is pure according to the orbital symmetries with Young scheme {43}, was constructed so that to describe, in the first place, the channel binding energy of the ground state as the p$^6$Li system and the mean square radius. Here, we slightly change its depth for the purpose to correctly describe the binding energy of $^7$Li in the n$^6$Li channel. In this case, the parameters of the pure $^2P^{\{43\}}$ Gaussian potential[85] of the n$^6$Li interaction for the GS of $^7$Li with $J^\pi = 3/2^-$ can be represented as

$$V_{GS} = -250.968085 \text{ MeV and } \gamma_{GS} = 0.25 \text{ fm}^{-2}. \tag{13}$$

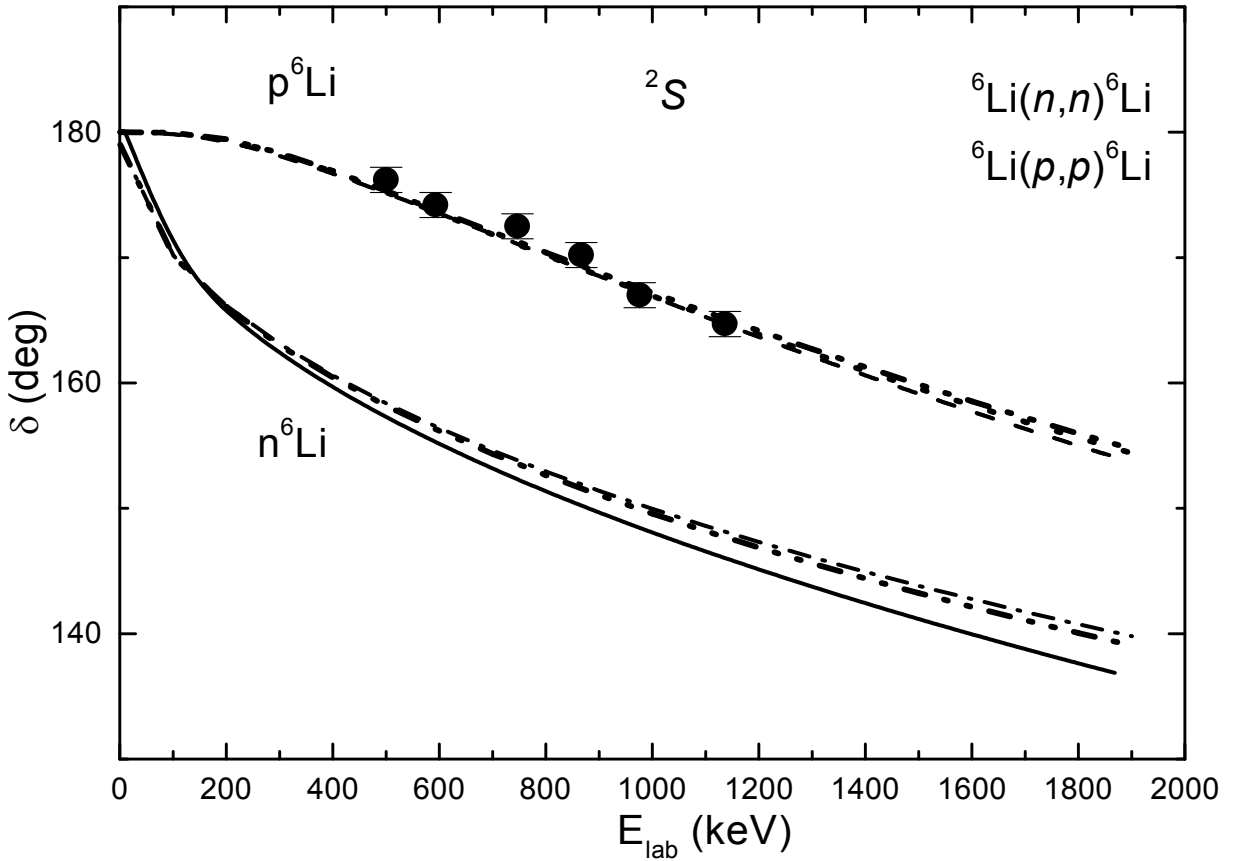

Fig. 6. The $^2S$ phase shifts of the n$^6$Li and p$^6$Li elastic scattering. Points: ● – extraction of the p$^6$Li phase shifts[28] from the experimental data.[29,30] Lines – the phase shifts of the elastic scattering with the potentials listed in the text.

The potential leads to the binding energy -7.249900 MeV at the experimental value -7.2499 MeV from Ref. 86 and has one forbidden state, corresponding to the Young scheme {61}.[3] The mean square charge radius is equal to 2.55 fm, which is on the whole in agreement with the experimental data,[86] where the value 2.39(3) fm is given. The neutron radius equals zero and the $^6$Li radius, which is slightly more than the $^7$Li radius and equals 2.51(10) fm,[86] are used for these calculations.

The value 2.45(1) was obtained for the asymptotic constant at the range 5÷15 fm. Let us note that according to Refs. 87, 88 where the different experimental data and theoretical results are collected, the value 1.76(14) fm$^{-1/2}$ is given, and after recalculation to the dimensionless value at $\sqrt{2k} = 1.05$ the value 1.68(13) was obtained. The value 1.890(13) fm$^{-1/2}$ is obtained in Ref. 89 that for dimensionless value leads to 1.80(1). This recalculation is necessary so far as the slightly different AC determination, which differs from our by the $\sqrt{2k}$ factor, was used in these works.

These potential parameters for the first excited state (FES) of $^7$Li with $J^\pi = 1/2^-$ were obtained:

$$V_{exc} = -248.935336 \text{ MeV and } \gamma_{exc} = 0.25 \text{ fm}^{-2}. \qquad (14)$$

The potential allows us to obtain the binding energy -6.772300 МэВ at the experimental value of -6.7723 МэВ,[86] the charge radius does not change relatively the previous results, and AC is equal to 2.33(1) at the interval 5÷15 fm. This potential has the forbidden bound state with the Young scheme {61}.

Another variant of the pure $^2P_{3/2}$ potential for n$^6$Li interaction of the GS of $^7$Li, but now without FS, can be represented as

$$V_{GS} = -75.190114 \text{ MeV and } \gamma_{GS} = 0.175 \text{ fm}^{-2}. \qquad (15)$$

It leads to the binding energy -7.249900 MeV and has only one bound allowed state, corresponded to the Young scheme {43}. The mean square charge and mass radii coincide with 2.54 fm, the AC at the range 5÷16 fm equals 2.03(1), that is only by 10÷15% differ from the results of Refs. 87-89.

The two-particle variational method with independent variation of parameters and with the expansion of the wave function by non-orthogonal Gaussian basis[61] was used for additional control of the correctness of calculations of the binding energy of $^7$Li in the GS potential with FS and AS. The binding energy -7.249898 MeV was obtained on the basis of this method at the dimension of the basis $N = 10$ and with independent variation of the parameters. The asymptotic constant $C_W$ of the variational WF remains at the level 2.45(5) at the range 5÷15 fm, and the residuals have the order of $10^{-11}$.[61]

Table 3. The variational parameters $\alpha_i$ and expansion coefficients $C_i$ of the GS WF of $^7$Li for the n$^6$Li channel.

| $i$ | $\alpha_i$ | $C_i$ |
|---|---|---|
| 1 | 2.468292899352664E-002 | -8.443780272416886E-004 |
| 2 | 5.659824615487678E-002 | -1.494186015886072E-002 |
| 3 | 1.229406461038807E-001 | -9.267494206256470E-002 |
| 4 | 2.513715488575826E-001 | -3.217760480847366E-001 |
| 5 | 7.328392817240388E-001 | 1.463594686074960 |
| 6 | 1.394554324801138 | 8.744682134317008E-001 |
| 7 | 1.968191404804425 | -2.564925474852117 |
| 8 | 2.224827222346167 | 3.963681316635119 |
| 9 | 2.494348228525606 | -2.317285290938208 |
| 10 | 2.835387525435829 | 485636531606636E-001 |

*Note.* Normalization coefficient of the wave function on the interval of 0÷25 fm is $N = 0.9999999999999947$.

As it was said before, the variational energy decreases with increasing of the basis dimension and yields the upper boundary of the true binding energy, and the finite-difference energy increases with decreasing step and increasing number of steps, then for the real binding energy of the n$^6$Li system it is reasonable to assume the average value for the binding energy -7.249899(1) MeV as valid. Thus, it may be considered that the error of determination of the binding energy using two different numerical methods (VM and FDM) and obtained on the basis of two different computer programs,[61] rewritten in Fortran-90[5] is on the level ±1.0 eV.

Completely identical results for binding energy, which is equal to -7.249900(1) MeV, i.e., obtained with error <±0.5 eV and other characteristics of the GS of $^7$Li in the n$^6$Li channel, are obtained for the GS potential without FS, the parameters of corresponding WF are listed in Table 4.

Table 4. The variational parameters $\alpha_i$ and expansion coefficients $C_i$ of the GS WF of the n$^6$Li system for the potential with one BS.

| $i$ | $\alpha_i$ | $C_i$ |
|---|---|---|
| 1 | 2.664737385927627E-002 | -1.095728292463283E-003 |
| 2 | 5.990159645362107E-002 | -1.553772918731152E-002 |
| 3 | 1.239933589538381E-001 | -8.044368539105093E-002 |
| 4 | 2.350818301848505E-001 | -2.013160349378503E-001 |
| 5 | 5.472422899975157E-001 | -1.115478059538508 |
| 6 | 6.087199937963733E-001 | 2.174880891838132 |
| 7 | 6.836630232030554E-001 | -1.459616783928601 |
| 8 | 8.216568512866885E-001 | 3.296229944619596E-001 |
| 9 | 1.245761137400640 | -2.961265316458718E-002 |
| 10 | 1.576324421758597 | 7.119580910569919E-003 |

*Note.* Normalization coefficient of the wave function on the interval of 0÷25 fm is $N = 1.000000000000001$.

### 3.2. *The total cross sections of the radiative neutron capture on $^6$Li*

During the consideration of the total cross sections of the radiative capture process we are taken into account the $E$1 transitions from the non-resonance $^2S$ and $^2D$ scattering states to the ground $^2P_{3/2}$ and first excited $^2P_{1/2}$ bound states of $^7$Li in the n$^6$Li channel. The calculation of the wave function of the $^2D$ wave without spin-orbital splitting was done on the basis of the $^2S$ potential at $L = 2$, but the accurate coefficients for the $E$1 transitions from the $^2D_{3/2}$ and $^2D_{5/2}$ scattering waves are taken into account in the expressions for the capture cross sections.[31,85] Consequently, it has turned out that the contribution of the $^2D$ scattering waves becomes appreciable only at the energies above 1.0÷1.5 MeV. In addition, in the real calculations only the GS potential was used for WF of both levels, since the WF of the ground and first excited states almost does not differ. Such assumption seems to be quite reasonable, so far as we consider only general form of total cross sections in the energy range with bounds difference other to eight orders. Here, we will not consider the capture process details, as it was done in review,[90] where the possibility of the description of the total cross sections of the photodisintegration process of $^7$Li into n$^6$Li channel at the energies 7.3÷8.8 MeV was considered in detail.

The potentials with two and one FS in the $^2S$ and $^2P$ waves, obtained from the

p$^6$Li scattering and checked earlier in the p$^6$Li system of $^7$Be, are used in the beginning of the consideration of the neutron capture on $^6$Li.[31,85] The calculation results for the total cross sections of the neutron capture on $^6$Li at the energies from 10$^{-5}$ to 3 10$^3$ keV for the first variant of the potentials with two BS in the $^2S$ and $^2P$ waves, are shown in Fig. 5 by the solid line, the dotted line is the results for the second combination of the potentials with one BS. The corresponding calculation results of the photodisintegration of $^7$Li in the n$^6$Li channel with the first variant of the potentials are shown in Fig. 4 by the solid line and for the second variant of the potentials – by the dotted one. It is seen from these figures that in both cases it is possible to obtain potentials, which describe the energy behavior of the total capture and photodisintegration cross sections at the energies from 25 meV to 1÷1.5 MeV completely correct. Such interactions are coordinated with the elastic scattering phase shifts and, in general, correctly describe some basic characteristics of the GS of $^7$Li, at that the variant of the potential without FS more correctly reproduces the AC value.

But if we will use the GS potential without FS, which describes the AC more exactly, for example, with the parameters

$$V_{GS} = -83.161074 \text{ MeV and } \gamma_{GS} = 0.2 \text{ fm}^{-2}, \quad (16)$$

then, using the $^2S$ scattering potential with one FS, we will obtain the result that is shown in Figs. 4 and 5 by the dashed line. Such potential leads to the value of binding energy of -7.249900 MeV, AC equals 1.85(1) at the range of 5÷13 fm, to the charge radius of 2.54 fm and mass radius of 2.53 fm.

The calculation results of the variational energy, which equals -7.249899 MeV, i.e., obtained with an accuracy of ±0.5 eV, and other characteristics of the GS of $^7$Li in the n$^6$Li channel for this potential are similar to the FDM results that was obtained above; the residuals have the order of 10$^{-10}$, the parameters of the WF are listed in Table 5.

Table 5. The variational parameters $\alpha_i$ and expansion coefficients $C_i$ of the GS WF of the n$^6$Li system.

| $i$ | $\alpha_i$ | $C_i$ |
|---|---|---|
| 1 | 2.665347013743804E-002 | -8.871735330500928E-004 |
| 2 | 5.940895728884596E-002 | -1.221361696531949E-002 |
| 3 | 1.219273413814190E-001 | -6.284879952239499E-002 |
| 4 | 2.340611751544998E-001 | -1.968287096274776E-001 |
| 5 | 4.751229388850844E-001 | -8.572931845080505E-001 |
| 6 | 5.485119023279393E-001 | 1.556074541398506 |
| 7 | 6.173563857857660E-001 | -1.203431194740232 |
| 8 | 7.395207514049224E-001 | 2.934610010474853E-001 |
| 9 | 1.003543127851490 | -3.090692233217297E-002 |
| 10 | 1.509188370554815 | 2.059998226181524E-003 |

*Note.* Normalization coefficient of the wave function on the interval of 0÷25 fm is $N = 0.9999999999999987$.

The calculation results of total cross sections for these potential that are shown in Fig. 5 by the dashed line also have a good agreement with the data[37-40] at 25 meV because of the big experimental errors, but lay slightly below the available data at the energy range from 100 keV÷1.0 MeV.[80-83] It is well seen from Fig. 4, that they probably lay between data from Ref. 81 and Ref. 82, which are shown by the black squares and open circles, respectively.

But, if the next parameters will be taken for the $^2S$ scattering potential:

$$V_S = -45.0 \text{ MeV and } \gamma_S = 0.25 \text{ fm}^{-2}, \qquad (17)$$

then the calculation results of total cross sections for the capture and photodisintegration cross sections, that are shown in Figs. 4 and 5 by the dashed-dot line, practically do not differ from the variant presented by the dotted line for the GS and scattering potentials with one BS. The phase shifts of such potential for both p$^6$Li and n$^6$Li scattering processes are shown in Fig. 6 by the dashed-dot-dot line. As it is seen from these results, it is possible to coordinate the description of the elastic scattering phase shifts and the basic characteristics of the GS of $^7$Li in the n$^6$Li channel, including radii and the AC value for the GS potential without FS.

The resonance 7.45 MeV Ref. 86 with the moment 5/2$^-$ (see Fig. 4), laying above the n$^6$Li threshold only by 0.2 MeV (c.m.), evidently applies to the $^4P_{5/2}$ scattering wave, and the possibility of the $M$1 transition during the photodisintegration of $^7$Li into the n$^6$Li channel, which is taken into account this state, was recently considered in review.[90] Let us note that this level, in principle, can be caused also by the $^2F_{5/2}$ scattering wave, although the existence of the resonance in the $F$ wave at so low energies seems to be doubtful.

Since at the energies from $10^{-5}$ and, about, to 100 keV the calculated cross section is almost straight line (see Fig. 5), it can be approximated by the simple function from Eq. (7) with the constant value 246.6118 μb·keV$^{1/2}$, which was determined by a single point at cross-sections with minimal energy of $10^{-5}$ keV. The absolute value of relative deviation of the calculated theoretical cross sections and the approximation of this cross section by the given above function from Eq. (8) in the range from $10^{-5}$ to 100 keV is less than 0.3%. If we will suggest that this form of the total cross section energy dependence will be preserved at lower energies, and we can perform an evaluation of the cross section value, for example, at the energy of 1 μeV ($10^{-6}$ eV=$10^{-9}$ keV), result is 7.8 b.

## 4.  Cluster n$^7$Li system

The knowledge of n$^7$Li interaction potentials in the continuous and discrete spectra is necessary for carrying out of the calculations of the total cross sections of the radiative neutron capture on $^7$Li at thermal and astrophysical energies. $^8$Li is stable nucleus, and in terms of strong interactions, since it decays towards $^8$Be only due to weak forces, it is reasonable to suppose that $^8$Li has two-cluster n$^7$Li structure and it is possible to use the known methods of the PCM with FS[3,16,59] for description of the corresponding characteristics. Let us note that, from the point of general view on this process and cluster structure of $^8$Li, the case of the n$^7$Li system with $LS$-coupling is considered here, but not the one when the neutron is in the $p_{3/2}$ state with the admixture of the $p_{1/2}$ relatively to $^7$Li, as it was done earlier in Refs. 91, 92 for the case of $jj$-coupling.

## 4.1. *The classification of cluster states in the n$^7$Li system*

Let us note, at first, that n$^7$Li system has the isospin projection $T_z = -1$, which is possible only in case of the total isospin $T = 1$.[93] Therefore, this system is pure in accordance with isospin, unlike to the p$^7$Li system that is mixed by isospin with $T = 0$ and 1,[93] so as for the p$^7$Be system at $T_z = +1$ and $T = 1$. At the same time, like in case of p$^7$Li system, the spin $S$, can be equal to 1 and 2, and some states of the n$^7$Li system also can be mixed by spin.[3]

Further, we will briefly stop on the classification by the orbital states of clusters of the treated system. It was shown in [11-13] that if the scheme {7} is used for $^7$Li, then possible Young schemes {8} and {71} in the 1+7 channel turn out to be forbidden, because of the rule indicates that there can not be more than four cells in a row.[94] They correspond to Pauli-forbidden states with relative motion moments $L = 0$ and 1, what is determined by Elliot rule.[94]

In the second case, when the orbital scheme of $^7$Li is equal to {43}, then the n$^7$Li, p$^7$Li or n$^7$Be, p$^7$Be systems contain forbidden states in $^3P$ waves with schemes {53} and in $^3S_1$ wave at the WF symmetry {44} and have the allowed $^3P$ state with the spatial scheme {431}. Thereby, the n$^7$Li potentials in the triplet spin state ought to have the forbidden bound $^3S_1$ state with the scheme {44} for scattering processes and the forbidden and allowed bound levels in $^3P$ waves with the Young schemes {53} and {431}, the last one is corresponded to the $^3P_2$ ground bound state of $^8$Li in the n$^7$Li channel. Further, we will consider thoroughly just that very case - the second variant of classifications of FS, using it, in the first place, for the GS of $^8$Li in the n$^7$Li channel.

The allowed symmetries at $S = 2$ and, consequently, bounded allowed levels in the n$^7$Li system are absent at any values of orbital moment $L$.[11-13] Thereby, the potential of the $^5S_2$ scattering wave has the bound FS with scheme {44}, and in $^5P$ scattering waves the potential has the FS with schemes {53} and {44}, at that the last of them can be in the continuous spectrum and the potential has only one bound FS with scheme {53}. This conclusion, evidently, is not unique and there is the version of $^5P$ scattering potentials with two bound FS for schemes {53} and {431}.

Maybe, as the third variant, it is possible to consider both allowed Young schemes {7} and {43} for the ground state of $^7$Li, since both of them are the FS and AS of this nucleus in the $^3$H$^4$He configuration.[11-13,16] This level classification will be slightly different, the number of FS will increase, and additional forbidden bound level will be added in each partial wave with $L = 0$ and 1.

But, since in the previous chapter for n$^6$Li system it was shown that it is possible to use the allowed scheme {42} without taking into account its forbidden configuration {6}, then further we will consider the second variant of the FS structure and potentials with the allowed scheme {43} for $^7$Li, as the main variant of the classification of FS and AS in such system. Therefore, we will consider that the potentials of the $^{3,5}S$ scattering waves, which are necessary for the consideration of the $E1$ electromagnetic transitions to the GS of $^8$Li at the neutron capture on $^7$Li, have the bound forbidden states with scheme {44}. The potential of the resonance $^5P_3$ scattering wave at 0.25 MeV, which allows to consider $M1$ transition to the GS of $^8$Li, can has one FS with scheme {53} or two bound forbidden states with {53} and {431}. The potential of the BS of $^8$Li in the n$^7$Li channel, which is the mixture of two $^3P_2$ and $^5P_2$ states, has one forbidden bound state with {53} and allowed bound state with {431} scheme, corresponding to the GS at the binding energy -2.03239 MэB.[93]

## 4.2. *Potential description of elastic n$^7$Li scattering*

We could not find any data on the elastic scattering phase shifts for the n$^7$Li or p$^7$Be systems at astrophysical energies.[35] Therefore, here we will construct the scattering potentials in the n$^7$Li system similar to the p$^7$Li scattering[11-13] and, on the basis of $^8$Li spectrum data,[93] given in Fig. 12 together with similar spectra of $^8$Be and $^8$B. The level spectra are shown so that to superimpose the level $2^+1$ of $^8$Li and $^8$B, which are the ground states, that stable for nuclear interactions and decay only due to weak forces.

The considered earlier bound state of the p$^7$Li system with $J^\pi T=0^+0$[93] corresponding to the ground state of $^8$Be, using the quantum addition rule of moments, could be formed only in the triplet spin state with $L=1$, so it is pure according to the $^3P_0$ spin state with $T=0$ (see Fig. 7).[93] Consequently, by the description of electromagnetic transitions, all previously obtained potentials for this system[3,6,11-13] are corresponded to the triplet spin state with the above obtained number of AS and FS. All electromagnetic transitions take place between different levels in triplet spin state, which has the allowed Young scheme and, consequently, the allowed bound state, corresponding to the ground state of $^8$Be in the p$^7$Li channel.

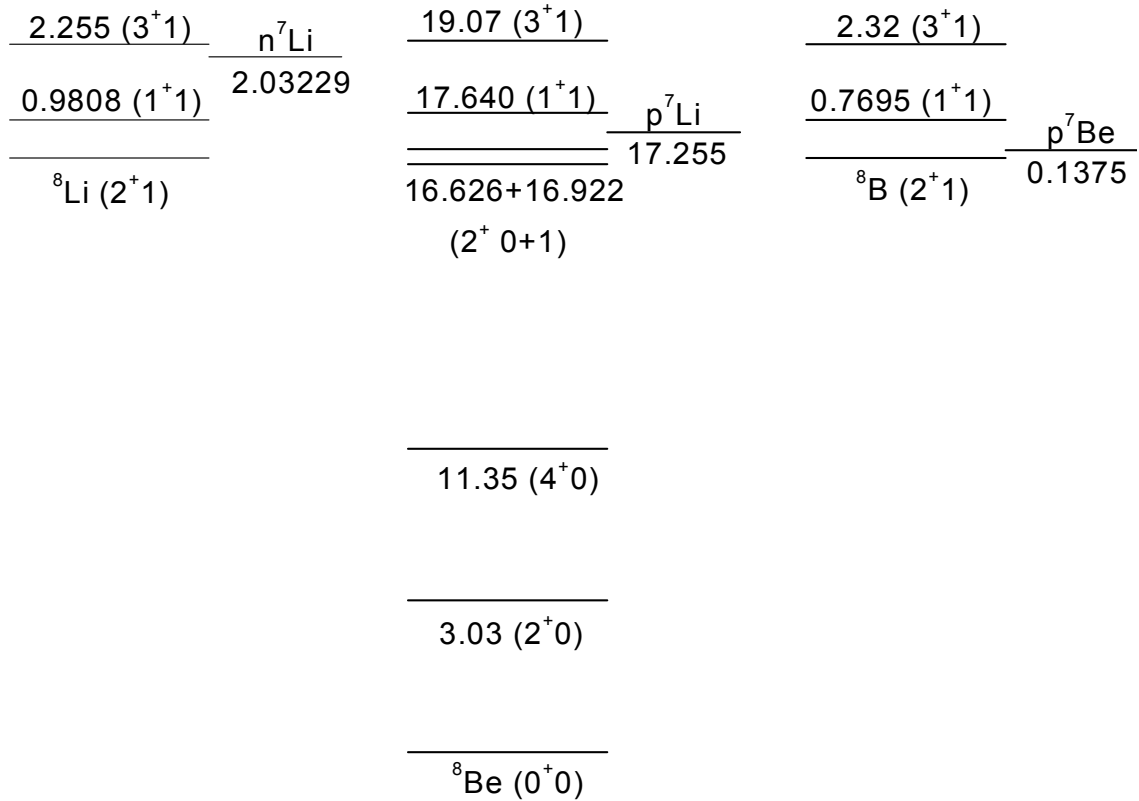

Fig.7. Energy level spectrum in MeV (c.m.) of $^8$Li, $^8$Be and $^8$B.[93]

Particularly, we have considered the $E1$ transition between the $^3S_1$ scattering state (mixed by isospin with $T=0$ and 1) to the ground bound $^3P_0$ state with $T=0$, and the $M1$ process between the resonance $^3P_1$ wave (with $T=1$) and the GS of $^8$Be, which have occurred with the isospin changes. Let us note that in case of p$^7$Li system with $T_z=0$, some scattering states, for example the $^3S_1$ wave, are mixed by isospin with $T = 0$ and 1. Therefore, practically the only part of the potential with $T = 1$ Refs. 3, 6, 11-13 was obtained for the $^3S_1$ wave, but the $^3P_1$ wave and its potential have the resonance of phase shift, i.e., the resonance level in $^8$Be, with experimentally obtained isospin T= 1. This state is pure by isospin. These two processes completely allow to describe the experimental data on the astrophysical *S*-factor of the radiative proton capture process on $^7$Li. It is considered that there is isospin changing for all these transitions, i.e., the rule $\Delta T=1$ is satisfied for them.[3,6,11-13]

In this case, the bound state of the n$^7$Li system with $J^\pi T=2^+1$, corresponding to the GS of $^8$Li, can be formed at $S=1$ and 2 with the orbital moment $L=1$, and is the mixture of the $^3P_2$ and $^5P_2$ states. In spite of absence of the allowed Young scheme at $S=2$, as it follows from the results of works[3,6,11-13] and from the given above classification, here it should be taken into account the presence in the GS of the $^5P_2$ wave admixture, that is necessary for consideration of the *M*1-transition from the $^5P_3$ resonance to the GS of $^8$Li.

The level $J^\pi T=3^+1$ in the $^8$Li spectrum (see Fig. 7) corresponds to the resonance $^5P_3$ phase shift of the n$^7$Li elastic scattering at the energy 0.22 MeV (c.m.) or 0.25 MeV (l.s.) above than n$^7$Li threshold.[93] This resonance $^5P_3$ state can be formed only with the total spin $S=2$, if to consider only minimally possible orbital moments, and is pure according to spin. Further, we will use the data for energy levels of $^8$Li and for widths of those levels[93] for construction of the potential corresponding to such resonance of the n$^7$Li scattering phase shift.

Certainly, the state $J^\pi T = 3^+1$ can be formed by the triplet $^3F_3$ configuration of the n$^7$Li system and the resonance will be present in the $^3F_3$ phase shift of the n$^7$Li elastic scattering. In this case, it is not necessary to assume the existence of admixture of the $^5P_2$ state in the GS of $^8$Li in the n$^7$Li channel, and it will be enough to consider only the $^3P_2$ configuration. But, one can draw a conclusion on the basis of all carried out phase shift analysis (see, for example, Refs. 52, 53) and analogous results for other similar cluster systems[3] that the presence of the resonance for the $^3F_3$ phase shift at so low scattering energy in the n$^7$Li system seems to be ambiguous.

The state with $J^\pi T = 1^+1$ that caused by $S = 1, 2$ and $L = 1$ is the $^{3+5}P_1$-level in the n$^7$Li channel is bounded at the energy 0.9808 MeV relative to the GS of $^8$Li or at the energy -1.05149 MeV relative to the n$^7$Li threshold.[93] Later we will also consider the *E*1 transitions to this level from the triplet and quintet *S* scattering waves. Therefore, all further results will concern to the $^7$Li(n, $\gamma_0$)$^8$Li and $^7$Li(n, $\gamma_1$)$^8$Li reactions and their cross sections. Further, by analogy with the p$^7$Li scattering and on the basis of data from Ref. 93, we will consider that $^3S_1$ and $^5S_2$ phase shifts in the range up to 1 MeV practically are equal to zero. It is confirmed by the absence of the resonance levels with negative parity in the spectrum of $^8$Li at such energies.

Since, earlier in the p$^7$Li system,[6,11-13,24-26] we have treated the variants of the potentials with two FS, then further, for comparison, we will also use potentials in all partial scattering waves with different number of FS that are required for calculations of the radiative capture. Initially we will find the *S*- and *P*-potentials with two FS, as it follows from the given above results, and then will consider the variants with one (the second classification variant) and zero FS, i.e., with their complete absence in each partial wave.

Practically zero phase shifts for the $^3S_1$ and $^5S_2$ waves of scattering at low energies can be obtained with parameters

$$V_S = -145.5 \text{ MeV and } \gamma_S = 0.15 \text{ fm}^{-2}. \qquad (18)$$

Here, we will consider such variant of the potential, since the similar potential was used while considering the p$^7$Li scattering in the $^3S_1$ state[6,11-13] - it contains two bound FS, as it follows from the classification of states for the third variant, at the bound FS with the orbital schemes {8} and {44}.

The zero phase shift can also be obtained with the potential

$$V_S = -50.5 \text{ MeV and } \gamma_S = 0.15 \text{ fm}^{-2}, \qquad (19)$$

which has only one bound FS for the second variant of classification with the scheme {44}, as well as with zero depth of the potential of Eq. (2) without FS, i.e., at the $V_0 = 0$ for both $S$ waves of scattering.

Certainly, the near-zero $S$ phase shifts can be obtained in both spin channels with the help of other variants of parameters of the Gaussian potential. In this case, it is impossible to fix parameters of such potential unambiguously, and the other combinations of $V_0$ and $\gamma$ with different number of FS are possible in the cases of Eqs. (18) and (19). However, as it will be shown further, the key role in the description of the total cross sections of radiative capture are played, evidently, not by the different combinations of parameters $V_0$ and $\gamma$, and not by the number of FS, but by the closeness to zero of scattering phase shifts obtained with these interactions.

The resonance $^5P_3$ phase shift of the $n^7$Li elastic scattering can be described by the Gaussian potential, for example, with parameters

$$V_P = -4967.45 \text{ MeV and } \gamma_P = 3.0 \text{ fm}^{-2}. \qquad (20)$$

This potential has two bound forbidden states, which can be compared to the schemes {53} and {431} for the second variant of classification FS, if to consider that the second FS with the scheme {431} is the bound state. The calculation results of the $^5P_3$ scattering phase shift are shown in Fig. 8 by the dotted line – the resonance is at the energy 254 keV (l.s.) with the width 37 keV (c.m.), and that is absolutely coincide with experimental value 254(3) keV.[93]

The parameters of the potential for one forbidden state with {53}, which is also corresponded to the second variant of classification on conditions that the FS with scheme {431} in the continuous spectrum, are represented as

$$V_P = -2059.75 \text{ MeV and } \gamma_P = 2.5 \text{ fm}^{-2}. \qquad (21)$$

The phase shift calculation results are shown in Fig. 8 by the dashed line – the resonance is obtained at the energy 254 keV. The width of the $^5P_3$ resonance is equal to 35 keV (c.m.) at the experimental values 35(5) or 33(6) keV (c.m.) according to different data from review.[93]

The potential parameters without FS, which will be considered additionally, are represented as

$$V_P = -425.1 \text{ MeV and } \gamma_P = 1.5 \text{ fm}^{-2}. \qquad (22)$$

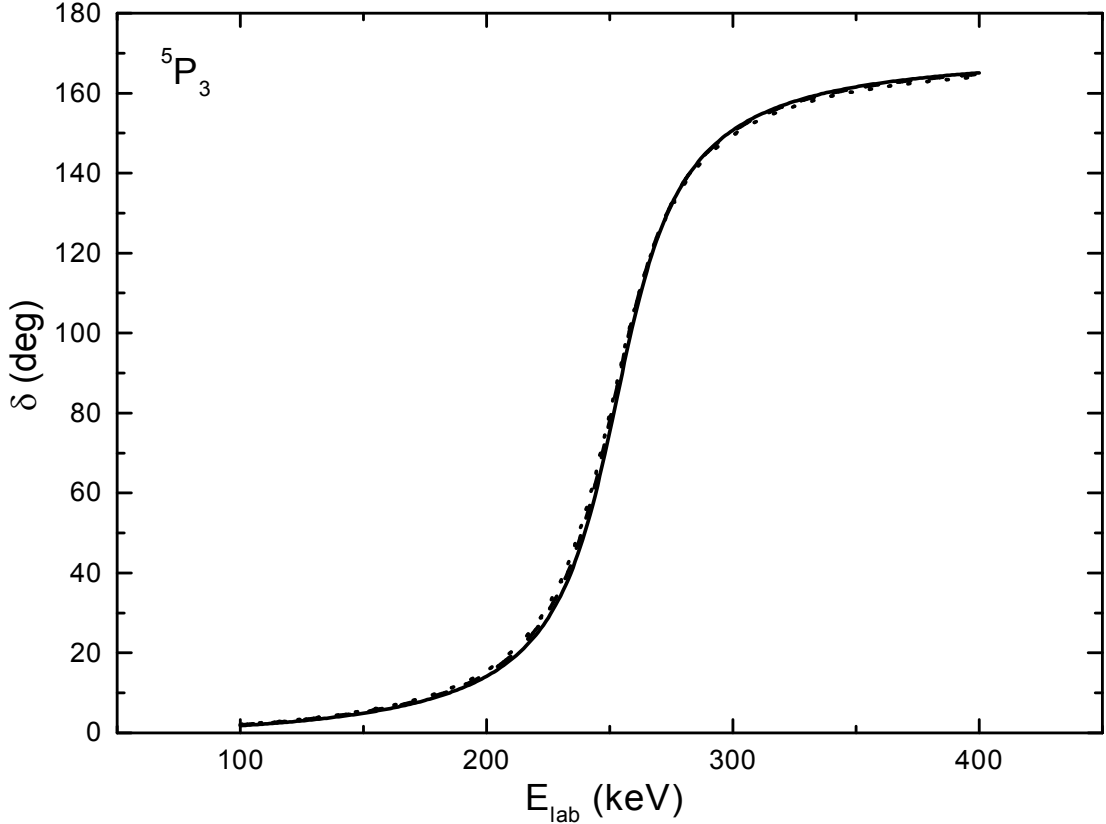

Fig. 8. The resonance $^5P_3$ phase shift of the n$^7$Li elastic scattering at low energies.

The calculation results of the $^5P_3$ phase shift with these parameters are shown in Fig. 8 by the solid line – the resonance is obtained at the energy 255 keV and its width is equal to 34 keV (c.m.). It should be stressed here, that the parameters of this potential at the given number of the bound FS are determined completely unambiguously according with the resonance energy and its width. The resonance width of the $^5P_3$ resonance is determined by the expression

$$\Gamma_{c.m.} = 2(d\delta/dE_{c.m.})^{-1}. \tag{23}$$

Since, we will consider the second variant of the cluster classification for the GS of $^8$Li then the following parameters for the potential for the bound $^{3+5}P_2$ state of the n$^7$Li system, which corresponds to the GS of $^8$Li in the considered cluster channel, can be used:

$$V_{GS} = -429.383779 \text{ MeV and } \gamma_{GS} = 0.5 \text{ fm}^{-2}. \tag{24}$$

Besides the allowed bound state corresponding to the GS of $^8$Li with {431}, such $^{3+5}P_2$ potential has the bound FS with {53}. The binding energy of -2.0322900 MeV that fully agrees with the experimental value,[93] the charge radius of 2.38 fm and the mass radius of 2.45 fm have been obtained with this potential on the basis of the finite-difference method (FDM)[61] with an accuracy $10^{-7}$ MeV. Apparently, the mean square charge radius of $^8$Li could not be substantially more than the $^7$Li radius, which is equal to 2.35(10) fm.[93] Therefore, the above obtained value of the mean square charge radius, in the n$^7$Li channel for the GS of $^8$Li, has the quite reasonable value. The zero value is used for the neutron charge radius, and its mass radius was taken equal to the corresponding radius of the proton 0.8775(51) fm.[95] The asymptotic constant for this GS potential was equal to $C_W = 0.78(1)$. The asymptotic constant error is determined by its averaging in the range of 4÷20 fm, where AC remains relatively stable.

Let us give, for comparison, the asymptotic constant of the n$^7$Li system $C(p_{3/2}) = 0.62$ fm$^{-1/2}$, obtained from the analysis of the experimental data in Ref. 92, which equals 0.81 after recomputation to the dimensionless value at $\sqrt{2k} = 0.767$. This value quite corresponds to the results obtained for the last variant of Eq. (18) of the GS potential of $^8$Li in the n$^7$Li channel. The value 0.78 fm$^{-1/2}$ is given in Ref. 96 that for dimensionless value leads to 1.02, in Refs. 87, 88 the value 0.74 fm$^{-1/2}$ was obtained and it leads to 0.96, and in Ref. 89 the value 0.59 fm$^{-1/2}$ was obtained for $C(^5P_2)$ and $C(^3P_2) = 0.28$ fm$^{-1/2}$, that in dimensionless form gives 0.77 and 0.36. Take notice of the fact that slightly differ AC determination was used in these works

$$\chi_L(R) = C_W W_{-\eta, L+1/2}(2k_0 R) \qquad (25)$$

and obtained their constants require recomputation.

The following parameters for the first excited state (FES) were obtained:

$$V_{ES} = -422.126824 \text{ MeV and } \gamma_{ES} = 0.5 \text{ fm}^{-2}. \qquad (26)$$

The allowed AS with {431} here corresponds to the first excited state of $^8$Li at 0.9808 MeV. In addition, this $^{3+5}P_1$ potential has the FS with {53} in full accordance with the second variant of the classification of orbital states. The binding energy -1.051490 MeV that fully agrees with the experimental value,[93] the charge radius 2.39 fm and the mass radius 2.52 fm have been obtained with this potential on the basis of the FDM[61] with an accuracy $10^{-6}$ MeV. The asymptotic constant for this potential was equal to $C_W = 0.59(1)$. The asymptotic constant error is determined by its averaging in the range of 4÷25 fm, where AC remains relatively stable.

For additional control of the bound energy calculations the two-particle variational method (VM) with the expansion of the cluster relative motion wave function for n$^7$Li system by non-orthogonal Gaussian basis with independent variation of parameters was used.[16,17] The energy -2.0322896 MeV, which was obtained at the dimension of the basis $N$=10 for the last variant of the GS potential of Eq. (26). The residuals have the order of $10^{-14}$,[61] the asymptotic constant at the range 5÷20 fm equals 0.78(1), the charge radius does not differ from the previous FDM results. Expansion parameters of the obtained variational GS radial wave function of $^8$Li in the n$^7$Li cluster channel are listed in Table 6.

Thereby, the average value of -2.0322898(2) MeV can be taken as realistic estimate of the binding energy in this potential. In other words, it may be considered that the accuracy of determination of the binding energy of $^8$Li in the n$^7$Li cluster channel for the GS potential from Eq. (24), using two numerical methods (FDM and VM) and based on two different computer programs, is at the rate of ± 0.2 eV.

The energy of -1.051488 MeV was obtained using VM for the first excited state with residuals of $10^{-14}$ – all other characteristics are not differ from the obtained above on the basis of the FDM. The expansion parameters of the WF is listed in Table 7, and the average energy can be written as -1.051489(1) MeV, i.e., the calculation error with such potential is equal to 1 eV and agrees with the FDM accuracy of $10^{-6}$ MeV.

Table 6. The coefficients and expansion parameters of the radial variational wave function of the ground state of $^8$Li for the n$^7$Li channel in non-orthogonal Gaussian basis.[16,17]

| $i$ | $\alpha_i$ | $C_i$ |
|---|---|---|
| 1 | 2.111922863906128E-001 | -1.327201117117602E-001 |
| 2 | 1.054889049037163E-001 | -4.625421860118692E-002 |
| 3 | 9.251179926861837E-003 | -1.875176301729967E-004 |
| 4 | 2.236449875501786E-002 | -2.434284188136483E-003 |
| 5 | 4.990617934603718E-002 | -1.282820835431680E-002 |
| 6 | 3.849142988488459E-001 | -2.613687472261875E-001 |
| 7 | 5.453825421384008E-001 | -2.108830320871615E-001 |
| 8 | 1.163891769476509 | 1.438162032150163 |
| 9 | 1.716851806191120 | 1.426517649534997 |
| 10 | 2.495389760080367 | 1.792643814712334E-001 |

*Note.* Normalization coefficient of the wave function on the interval of 0÷25 fm is $N = 9.999998392172028E-001$.

Table 7. The coefficients and expansion parameters of the radial variational wave function of the first exited state of $^8$Li for the n$^7$Li channel in non-orthogonal Gaussian basis.[61]

| $i$ | $\alpha_i$ | $C_i$ |
|---|---|---|
| 1 | 2.034869839899546E-001 | -1.268995424220545E-001 |
| 2 | 9.605255016688968E-002 | -4.250984818616291E-002 |
| 3 | 6.473027608029138E-003 | -2.029700124120304E-004 |
| 4 | 1.743880699865412E-002 | -2.308434897721290E-003 |
| 5 | 4.241481028548091E-002 | -1.167539819061673E-002 |
| 6 | 3.943411589808715E-001 | -2.876208138367455E-001 |
| 7 | 5.758070107927670E-001 | -1.307197681388061E-001 |
| 8 | 1.148526246366072 | 1.335023264621784 |
| 9 | 1.706295940575450 | 1.303208908841006 |
| 10 | 2.491484117851039 | 1.558051077479201E-001 |

*Note.* The normalization factor of the wave function on intervals 0÷25 Fm is equal $N = 9.999907842436313E-001$.

### 4.3. *The $^7L(n,\gamma)^8Li$ radiative capture*

Let us note that, evidently, the $E$1-transition in the n$^7$Li system, for the first time, was considered in Refs. 97, 98, where possibility of the correct description of the total cross sections in non-resonance energy range was shown on the basis of one particle model with the Woods-Saxon potential agreed with energy levels of $^8$Li. Later on, this process was considered on the basis of direct capture model, for example, in Ref. 91. Similar results concern to Refs. 18-23, where there is acceptable description of the total capture cross sections on the basis of the $E$1-process, but without taking into account their resonance behavior.

As far as we know, the results with an acceptable description of these total radiative capture cross sections in the resonance energy range at 0.25 MeV were obtained only recently,[99] on the basis of model-independent methods. Further, we will show that similar results of describing this resonance on the basis of the $M1$ transition from the $^5P_3$ scattering wave, which has the resonance at this energy, to the $^5P_2$ component of the GS WF of $^8$Li in the n$^7$Li channel can be realized in the potential cluster model.

We will take into account the $E1$ transition from the non-resonance $^3S_1$ scattering wave to the triplet $^3P_2$ part of the GS WF during the consideration of the electromagnetic processes in the $^7$Li(n,γ$_0$)$^8$Li reaction and, as before, for the proton capture on $^7$Li.[11-13] In addition, in comparison with the p$^7$Li system, the transition from the quintet $^5S_2$ scattering wave to the quintet $^5P_2$ part of wave function of the GS of $^8$Li will be added. And, as we have said before, the $M1$ transition from the resonance $^5P_3$ wave with the level at $J^\pi T = 3^+1$ (see Fig. 8) to the quintet $^5P_2$ part of the GS WF will be taken into account additionally. The $E1$ process from both $^{3+5}S$ scattering waves to the first excited $^{3+5}P_1$ state is taken into account.

Thus, the total cross section of the capture process, taking into account of all considered here electromagnetic transitions for neutron capture on $^7$Li, can be represented in the form:

$$\sigma(E1+M1) = \sigma(E1,{}^3S_1 \to {}^3P_2) + \sigma(E1,{}^5S_2 \to {}^5P_2) + \sigma(M1,{}^5P_3 \to {}^5P_2) \quad (27)$$

and

$$\sigma_1(E1) = \sigma(E1,{}^3S_1 \to {}^3P_1) + \sigma(E1,{}^5S_2 \to {}^5P_1) \quad (28)$$

In the frame of considered model there is no possibility to extract the $^5P_2$ and $^3P_2$ parts in the GS WF, so we will use the mixed by spin $P_2$ function that is obtained with the BS potential, for example, from Eq. (24). The results of carried out calculations are compared with the experimental measurements of the total cross sections of the capture reaction at the energy range from 5 meV to 1.0 MeV, which were done in Refs. 49 ,50 ,100-104.

Since, three variants of the potentials for each partial scattering wave were given; we are not going to explain the results for each of these combinations in detail. At once, we will give the final and, apparently, the best result for the calculation of the total cross section for the radiative neutron capture on $^7$Li to the GS at the energy up to 1 MeV (l.s.), which is shown by the dot-dashed line in Fig. 14. These results were obtained for the potential of the GS from Eq. (24), for the $S$ scattering waves in triplet and quintet states with parameters from Eq. (19) and for the potential of the $^5P_3$ resonance scattering wave with parameters Eq. (21). The dashed line shows the cross section corresponding to the sum of the $E1$ transitions from the $^3S_1$ and $^5S_2$ waves to the GS, dotted line – the cross section of the $M1$ transition between the $^5P_3$ scattering state and the GS of $^8$Li in the n$^7$Li channel. The results for total cross sections, which take into account all processes at the GS and FES, were shown by the solid line. The potential from the Eq. (26) and the same potentials of the $^{3+5}S$ scattering waves are used for the FES.

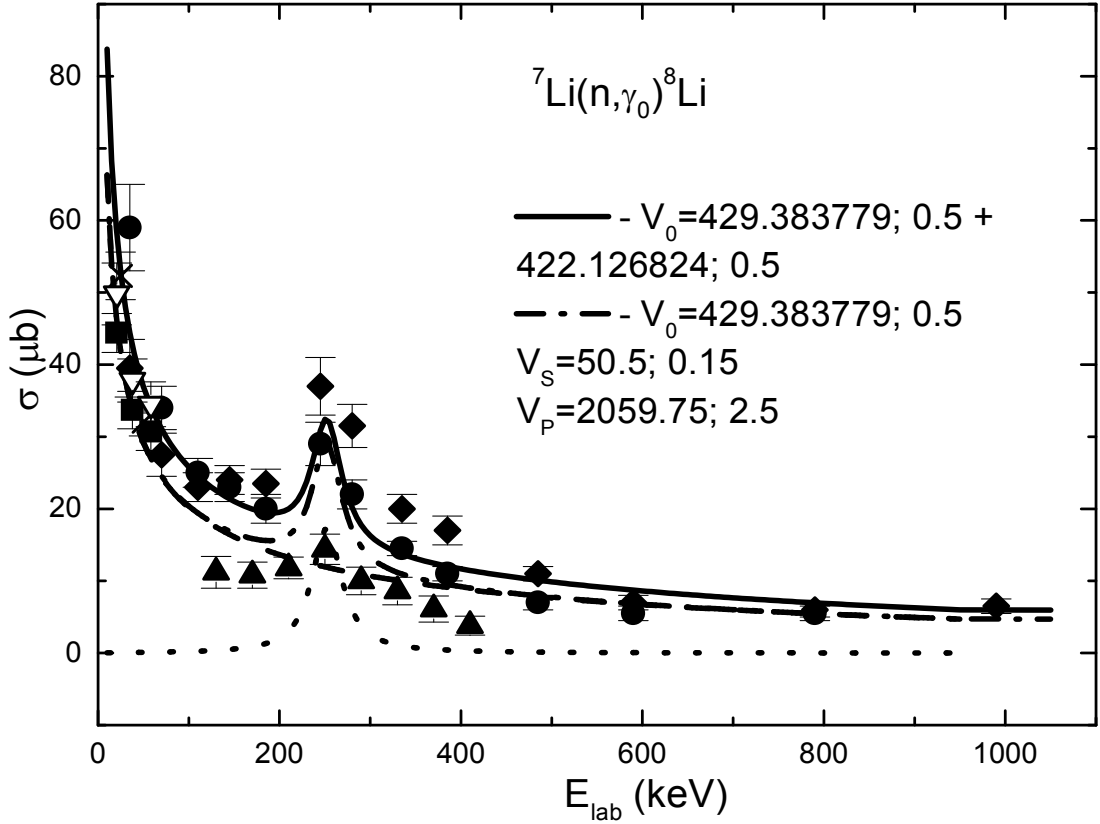

Fig. 9. The total cross sections for the radiative neutron capture on $^7$Li. The experimental data: ● and ♦ from Ref. 100, ■ – Ref. 101 for capture to the GS and ∇ - summarized cross sections for capture to the GS and FES, ▲ – Ref. 102, x – Refs. 49, 50. Lines: results of the calculation for different electromagnetic transitions with the potentials listed in the text.

More detailed shape and value of the calculated total cross sections for these variant of the potentials at energies from 5 meV to 150 keV are shown in Fig. 15. It is clear from these results that using the scattering potentials with one FS in the $S$ and $P$ waves, obtained in the frame of the potential cluster model, it is possible to describe the existent experimental data in the widest energy range - from 5 meV to 1.0 MeV. It should be noted that the results, shown in Fig. 15 by the open circles, are the cross sections that were measured only for the capture to the GS of $^8$Li in Ref. 103. The measurements for capture to the GS (black squares) so as for the total cross sections taking into account transitions to the GS and FES (reverse open triangles) were done in Ref. 101.

The usage of the variant of scattering potential for the $^5P_3$ wave with two FS practically does not change the results of description of the resonance at 0.25 keV. Thereby, the noted above ambiguity of the FS number does not influence to the results. The calculation of the cross sections with the interaction without FS leads to the appreciable decrease of the cross section value at the resonance energy, i.e., to the worsening of quality of the cross section description in this energy range. The variants of scattering potentials in the $S$ waves with two FS or without FS almost do not change the cross section calculation results. Even the double change of the width of $S$ potentials, notably to 0.3 fm$^{-2}$ at the depth 100 MeV, weakly affects the value of the calculated cross sections. Only the close to zero values of the scattering phase shifts at spin $S = 1$ and 2 are significant for these partial waves.

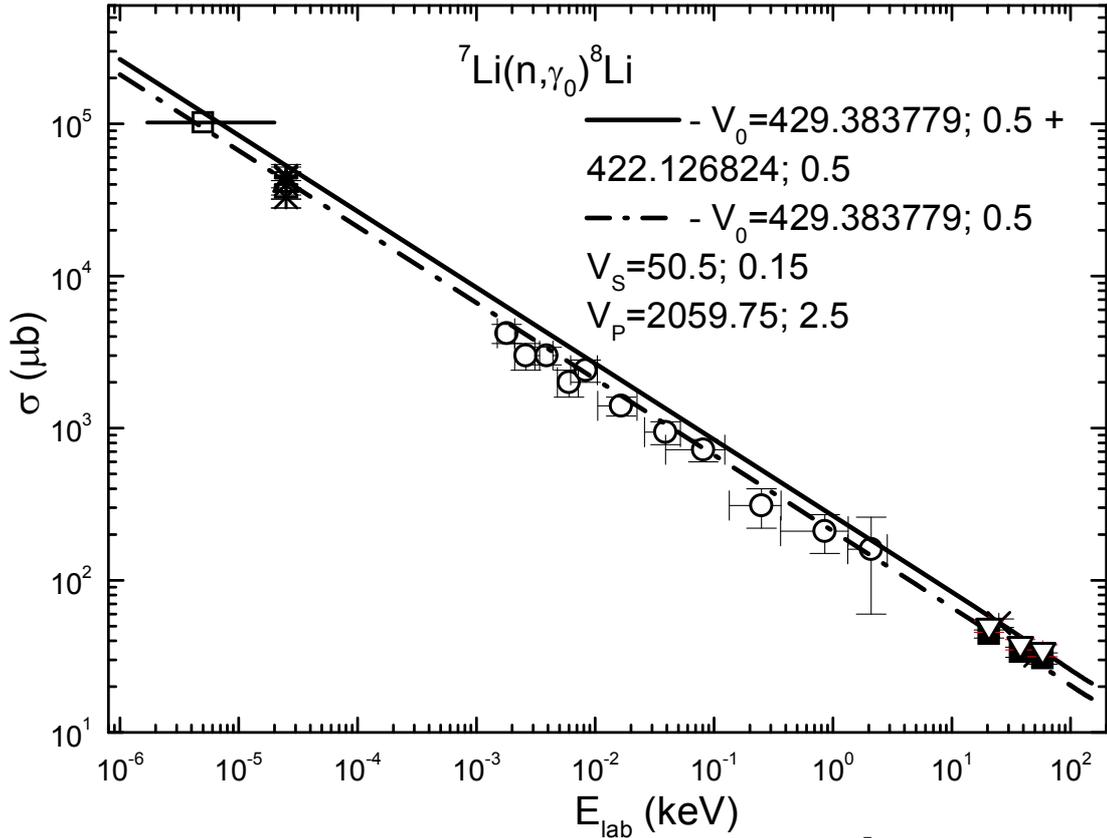

Fig. 10. The total cross sections for the radiative neutron capture on $^7$Li at ultralow energies. Experimental data: ■ – Ref. 101 for capture to the GS and ▽ - summarized cross sections for capture to the GS and FES, ○ – Ref. 103, □ – Refs. 18-23, the solid horizontal line - energy range where the measurements were done,$^{18\text{-}23}$ △ – Ref. 100, * – results of Ref. 104 and other data given there, x – Refs. 49, 50. Lines: as in Fig. 9.

Consequently, we can think that only the ground state potential of $^8$Li in the n$^7$Li channel with one FS from Eq. (24) and corresponded scattering potentials from Eqs. (19) and (21) leads to a reasonable description of the available experimental data of total cross sections of radiative capture process for all considered energy region, with bounds difference other to nine orders. The presence or absence of the FS in the potentials of $S$ scattering waves does not play any role, only zero (0°±2°) values of the scattering phase shifts are important. Small changes of the value of the calculated total cross sections does not allow to draw definite conclusions with respect to the number of the FS for the $^5P_3$ scattering potential in the resonance energy range.

Similarly to the previous systems and since at the energies from 1 meV to 100 keV the calculated cross section is almost a straight line (see Fig. 10, solid line), it can be approximated by the simple function from Eq. (7). The constant value of 265.7381 μb·keV$^{1/2}$ was determined by a single point at cross-sections with minimal energy of 1 meV. As in the previous cases, it is possible to consider the absolute value of relative deviation of the calculated theoretical cross sections and the approximation of this cross section by the expression Eq. (7) as a function of energy in the range from 10$^{-6}$ to 100 keV. Then we will obtain that at the energies lower than 100 keV this deviation is at the level of 1.0%. If, as usual, we will suggest that this form of the total cross section energy dependence Eq. (7) will be preserved at lower energies, and we can perform an evaluation of the cross section value, for example, at the energy of 1 μeV (10$^{-6}$ eV=10$^{-9}$ keV), result is 8.4 b. The cross section approximation coefficient for dashed-dot line in Fig. 15 is equal to 210.538 μb·keV$^{1/2}$, and the cross section value at 1 μeV equals 6.7 b.

## 5. The radiative neutron capture on $^{12}$C and $^{13}$C

Turning to the study of heavier nuclei in terms of the same cluster model, let us consider the possibility to describe the experimental data on total cross sections of the radiative neutron capture on $^{12}$C and $^{13}$C at the energy range from 25 meV to 1.0 MeV. Thereto, we have taken into account not only the $E$1 transition from the certain scattering states to the ground state of $^{13}$C and $^{14}$C in the n$^{12}$C and n$^{13}$C channels, but the capture on the three low laying excited states 1/2$^+$, 3/2$^-$ and 5/2$^+$ of $^{13}$C was calculated.

### 5.1. *Total cross sections for the neutron capture on $^{12}$C*

The $E$1($L$) transition, which is caused by the orbital part of the electric operator $Q_{JM}(L)$,[16] is taken into account in the calculations of the process of radiative neutron capture on $^{12}$C. Such transition in the n$^{12}$C → $^{13}$Cγ process is possible between the doublet $^2S_{1/2}$ scattering state and the ground bound $^2P_{1/2}$ state of $^{13}$C in the n$^{12}$C channel. At that, here we are considering not only the $E$1 transition to the ground state of $^{13}$C, but the capture on the three low laying excited states 1/2$^+$, 3/2$^-$ and 5/2$^+$ of $^{13}$C was calculated too. The two-particles interaction potentials as usually are constructed on the basis of the elastic scattering phase shifts and on the acceptable description of the basic characteristics of the BS of $^{13}$C in the cluster channel.[3]

The classification of the orbital states for n$^{12}$C and p$^{12}$C systems by Young schemes was treated in Ref. 51. It was shown that complete system of 13 nucleons has the next set of Young schemes {1}×{444} = {544} + {4441}.[105] The first of the obtained scheme is compatible with the orbital momentum $L = 0$ and is forbidden, so far as it could not be five nucleons in the $s$-shell. The second scheme is allowed and compatible with the angular moments $L = 1$ and 3 defined according the Elliot rules.[105] State with $L = 1$ corresponds to the ground bound allowed state of $^{13}$C in the n$^{12}$C channel with quantum numbers $J^\pi$, $T = 1/2^-$, 1/2. So, there might be one forbidden bound state in $^2S$ wave potential, and $^2P$ wave should have the allowed state only in the n$^{12}$C channel at the energy -4.94635 MeV.[106]

However, we regard the results on the classification of $^{13}$C and $^{14}$C nuclei by orbital symmetry in n$^{12}$C and n$^{13}$C channels as the qualitative ones as there are no complete tables of Young schemes productions for the systems with a number of nucleons more than eight,[107] which have been used in earlier similar calculations.[3,16] At the same time, just on the basis of such classification we succeeded with description of the available experimental data on the radiative proton capture on $^{12}$C.[51] Therefore, during the consideration of the n$^{12}$C system, we will use here the given above classification of the orbital states, which leads us to the definite number of FS and AS in the interaction potentials that allows us to fix their depth quite definitely.

The interaction potential for $^2S_{1/2}$ wave of the p$^{12}$C scattering was constructed earlier in Refs. 3-6, 51 in a way to describe correctly the corresponding partial elastic scattering phase shift, which has the pronounced resonance at 0.42 MeV. The n$^{12}$C system, considered here, has no resonances at energies up to 1.9 MeV according review Ref. 106. So, its $^2S_{1/2}$ phase shift should reveal relatively smooth behavior in this energy region. We were unable to find in the literature the results of the phase shift analysis for n$^{12}$C elastic scattering at the energies below 1.0÷1.5 MeV,[35,36] although its results should differ notably from the analogous for p$^{12}$C scattering.[51]

That is why for determine of the proper behavior of the $^2S_{1/2}$ phase shift, which is required for previous calculations, the phase shift analysis of the n$^{12}$C elastic scattering was done at astrophysical energies, viz. from 50 keV to 1.0 MeV.[108] The experimental measurements of differential elastic scattering cross sections in the energy range from 0.05 up to 2.3 MeV was done in Ref. 109. The results of our analysis for the $^2S_{1/2}$ phase shift are presented in Fig. 11a by black points. Let us note that, since the $^2S_{1/2}$ phase shift has the forbidden bound state, its values in Fig. 11a start from 180°.[14]

Let us proceed now to the description of the results for the GS potential, and then we will return to the phase shift analysis and potentials of scattering processes. The potential, for the ground state of $^{13}$C in the n$^{12}$C channel for the $^2P_{1/2}$ wave without FS, was constructed following the results obtained earlier for p$^{12}$C system.[51] This potential should correctly reproduce the binding energy of $^{13}$C in the n$^{12}$C channel equals -4.94635 MeV,[106] as well as the value of mean square radius of $^{13}$C, which is equal to 2.4628(39) fm.[106] The value 2.472(15) fm from Ref. 110 was taken for charge and mass radius of $^{12}$C; the neutron charge radius is zero, and its mass radius is taken as proton one 0.877(5) fm.[95]

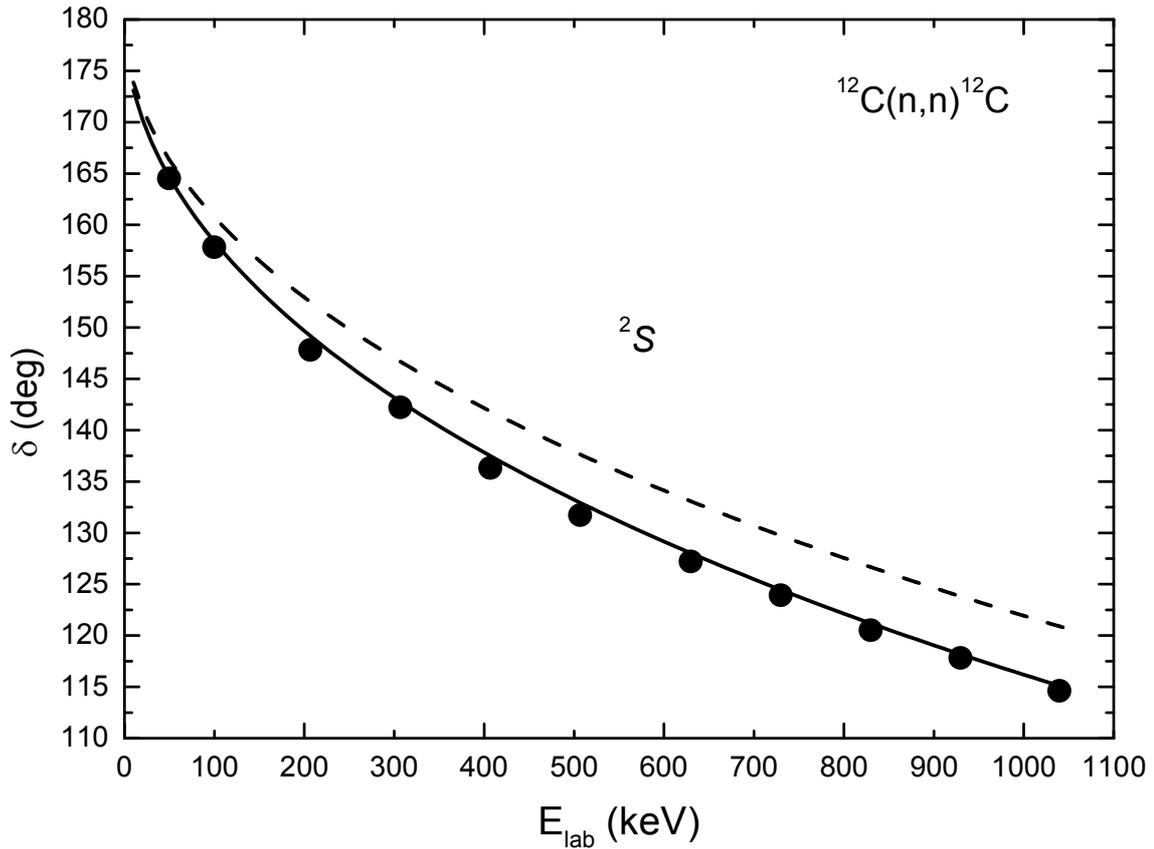

Fig. 11a. The $^2S_{1/2}$ phase shift of the n$^{12}$C scattering at low energies. The results of our phase shift analysis for the $^2S$ phase shift are given by black points (●) from Ref. 108. Lines – the calculations with different potentials that are given in text.

Using the previous results for the p$^{12}$C channel of $^{13}$N, the following parameters for considered here n$^{12}$C system were obtained

$$V_{GS} = -135.685683 \text{ MeV}, \quad \gamma_{GS} = 0.425 \text{ fm}^{-2}. \tag{29}$$

The potential gives the binding energy -4.946350 MeV with accuracy $10^{-6}$ MeV by FDM, mean square charge radius $R_{ch}$ = 2.48 fm, and mass radius $R_m$ = 2.46 fm. The asymptotic constant turned equals 0.99(1) on the interval 5÷13 fm. The given error for the AC is defined by averaging over the pointed interval of distances.

Let us note, that according data from Refs. 87, 88, where the compilation of many results is presented, the obtained value for this constant recalculated with $\sqrt{2k} = 0.942$ to the dimensionless quantity is 1.63(4). According data from Ref. 22, this value after recalculation is equal to 2.05(18). Let's take note, that such recalculation is coming due to another specification for AC differing from our by factor $\sqrt{2k}$.

There is another variant of the n$^{12}$C potential reproducing the ground state of $^{13}$C, with parameters:

$$V_{GS} = -72.173484 \text{ MeV}, \quad \gamma_{GS} = 0.2 \text{ fm}^{-2}. \quad (30)$$

This potential leads to the binding energy -4.94635034 MeV with accuracy $10^{-8}$ by FDM and same charge radius 2.48 fm, but mean square mass radius $R_m$ = 2.50 fm is a little bit greater, and AC equals 1.52(1) within the interval 5÷18 fm agrees better with data in Ref. 22, 87, 88. Solid line in Fig. 11b shows WF of such $^2P_{1/2}$ potential.

The variational method was applied as additional computing control for the calculation of binding energy.[61] It gave the energy value -4.94635032 MeV with dimension $N = 10$ and independent parameter varying of the GS potential of Eq. (30). The asymptotic constant $C_W$ of the variational WF, which parameters are listed in Table 8, is 1.52(2) within the interval 5÷15 fm while the residual does not exceed $10^{-12}$.[61] The charge radius is the same as obtained by FDM.

Table 8. The variational parameters and expansion coefficients of the radial WF of the GS of $^{13}$C in the n$^{12}$C channel for the potential of Eq. (30).

| $i$ | $\alpha_i$ | $C_i$ |
| --- | --- | --- |
| 1 | 1.500426018861289E-002 | 1.223469853688857E-004 |
| 2 | 1.002841633851088E-001 | 3.503273917493124E-002 |
| 3 | 1.981842450457470E-001 | 1.115174300485543E-001 |
| 4 | 3.011361231511710E-002 | 1.898077834207565E-003 |
| 5 | 1.460253375610869E-001 | 2.604340601242970E-002 |
| 6 | 5.115290090973104E-001 | 9.245769209919236E-002 |
| 7 | 9.742057085044215E-001 | -2.382087077902581E-003 |
| 8 | 3.220854607507809E-001 | 1.870518591470587E-001 |
| 9 | 8.801958230927104E-001 | 7.197537136787223E-003 |
| 10 | 5.612447142811238E-002 | 1.050288601397638E-002 |

*Note.* The normalization of the function in the range 0÷25 fm equals $N = 0.9999999999697765$.

As the variational energy decreases at the increasing of basis dimension and reaches the upper limit of true binding energy, and the finite-differential energy increases at the reducing of step value and increasing of step number, then it is reasonable to assume the average value for the binding energy -4.94635033(1) MeV as valid. Consequently, the accuracy of determination of binding energy, based on two different methods and calculated on the basis of two different computer programs, is on the level $\pm 10 \cdot 10^{-9}$ MeV = $\pm 10$ meV.

Following the declared isobar-analogue concept, the potential for $^2S_{1/2}$-wave of $n^{12}C$-scattering with parameters obtained for $p^{12}C$-scattering

$$V_S = -102.05 \text{ MeV}, \quad \gamma_S = 0.195 \text{ fm}^{-2}, \qquad (31)$$

which do not lead to the resonance as it is shown by the dashed curve in Fig. 11a, if the Coulomb potential is switched off. It gives the total radiative capture cross sections several orders less comparing the experimental data within the treated energy range from 25 meV up to 1 MeV.

Let us turn to the potential describing well $^2S$ phase shift of the $n^{12}C$ elastic scattering with parameters in Ref. 108

$$V_S = -98.57558 \text{ MeV}, \quad \gamma_S = 0.2 \text{ fm}^{-2}. \qquad (32)$$

Phase shift corresponding this potential, is given by solid curve in Fig. 11a. Phase shift of this potential is given by solid curve in Fig. 11a, and dashed curve in Fig. 11b shows corresponding WF. Dots on Fig. 11b display the integrand of $I_J$ overlapping integral corresponding to $E1$ transition as the product of $^2S_1$ scattering wave in $n^{12}C$ channel, $^2P_{1/2}$ ground state cluster WF of $^{13}C$, and relative distance r.

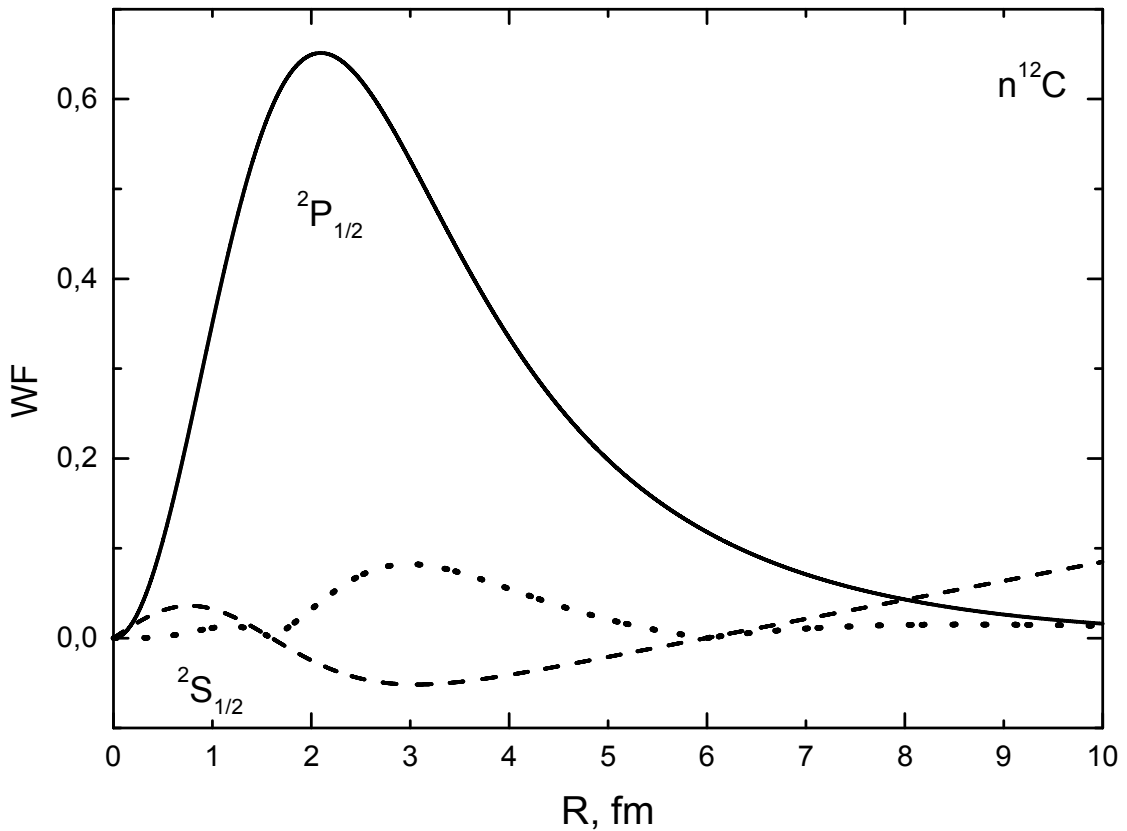

Fig. 1b. Radial wave functions of $^2P_{1/2}$ ground state of $^{13}C$ in the $n^{12}C$ channel and $^2S_{1/2}$ scattering wave at 10 keV. Dots display the integrand of $I_J$ for $E1$ transition.

Note, if we switch off the Coulomb interaction in initial $^2S_{1/2}$ potential determined for the $p^{12}C$ scattering for the correct reproducing of subthreshold resonance at 0.42 MeV this state becomes bound. So, the potential in $n^{12}C$ channel besides the forbidden state has now allowed one corresponding to the first excited state (ES) of $^{13}$C at 3.089 MeV with $J^\pi = 1/2^+$. Therefore for the correct description of binding energy in $^2S_{1/2}$ wave laying at -1.856907 MeV towards the threshold of n$^{12}$C channel parameters of potential of Eq. (32) are given with high accuracy.

Total cross section obtained with Eq. (29) for BS and scattering potential of Eq. (32) is shown by dashed curve in Fig. 12a. Calculated cross section is factor two lower than experimental data at 25 meV,[43] and it lies a little lower than data from Refs. 47, 49 in the energy range 20-200 keV. For comparison results with the same scattering potential of Eq. (32), but binding ground state potential of Eq. (30) are given in Fig.12a by solid line. They lead to somewhat a compromise with different experimental data at 25 meV,[43] but in the energy range 20-200 keV shift a little to the average values obtained in Refs. 45-49.

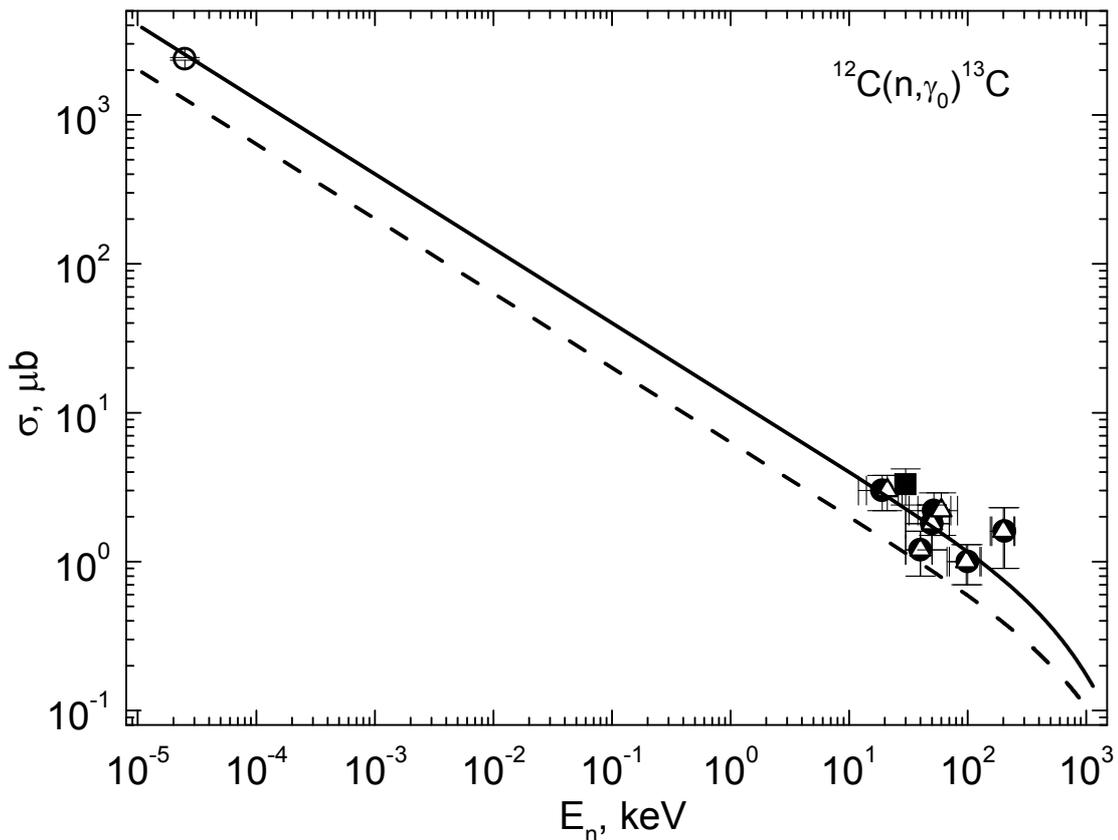

**Fig. 12a.** Total cross sections of the radiative neutron capture on $^{12}$C at low energies. Experimental data: black squares (■) – Ref. 48, black dots (●) – Ref. 49, open triangles (Δ) – Ref. 47, and open circles (○) – Ref. 43. Solid line – total cross section calculated with GS potential of Eq. (29) and scattering potential of Eq. (32); dashed curve – with GS from Eq. (30) and scattering potential of Eq. (32).

We would like to emphasize that these results have been obtained for the potentials of Eqs. (32) and (30) conformed GS characteristics of $^{13}$C, viz. asymptotic constant and low energy n$^{12}$C elastic scattering phase shifts. Thereby, this combination of potentials, describing the characteristics of both discrete and continuous spectra of n$^{12}$C system, allows to reproduce well available experimental data on the radiative neutron capture cross sections for transitions to the GS in the energy range from 25 meV up to 100÷200 keV covering seven orders.

Now treating transitions on to exciting states we want to remark that AC given in Ref. 96 for the first ES 1/2$^+$ of $^{13}$C in n$^{12}$C channel is 1.61 fm$^{-1/2}$, and recalculated dimensionless value turned to be 1.66. Besides, in Refs. 18-21 AC equals 1.84(16) fm$^{-1/2}$ was obtained for the first excited state, or recalculated value 1.90(17).

In the present case E1 transition from $P_{1/2}$ and $P_{3/2}$ scattering waves onto $S_{1/2}$ binding excited state in n$^{12}$C channel with Eq. (32) is assumed. As $P$ wave has no FS, and there are no negative parity resonances in spectrum of $^{13}$C nucleus then corresponding potentials may be regarded zero. While constructing a potential for the binding ES we would orient on the reproducing of the mentioned AC value, as its width affects weakly on the mean square radius.

As a result, for excited BS in $S_{1/2}$ wave with FS potential of Eq. (32) was used. It leads to the binding energy -1.856907 MeV with accuracy $10^{-6}$ by FDM, charge radius 2.48 fm, mass radius 2.67 fm, and AC equals 2.11(1) within the interval 6-24 fm. AC values do not differ too much from results of Refs. 18-21. Total cross sections, given on Fig. 12b by solid line, display reasonable agreement with experimental data at low energies.

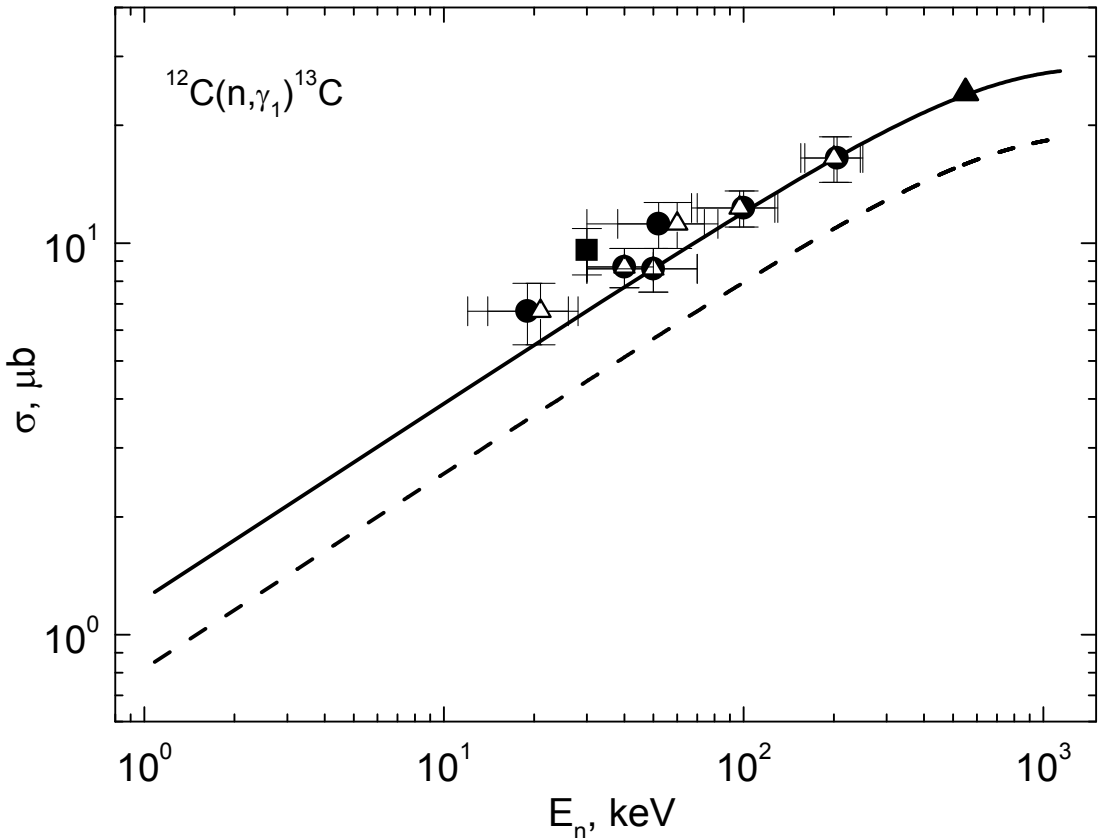

Fig. 2b. Total cross sections of the radiative neutron capture on $^{12}$C on the first excited state 1/2$^{+}$ of $^{13}$C at low energies. Experimental data: black squares (■) – Ref. 48, black dots (●) – Ref. 49, open triangles (Δ) – Ref. 47, closed triangles ▲ – Ref. 45. Solid line – total cross section calculated with ES potential of Eq. (32) and scattering $P$ potential with zero depth; dashed curve – with ES from Eq. (33) and scattering $P$ potential with zero depth

Note, more shallow potential may be used for this excited state, for example,

$$V_S = -184.84634 \text{ MeV and } \gamma_S = 0.4 \text{ fm}^{-2}, \tag{33}$$

It leads to the binding energy -1.856907 MeV with accuracy $10^{-6}$, charge radius 2.49 fm, mass radius 2.60 fm, and AC equals 1.71(1) within the interval 5-25 fm. This value of AC is more consistent with results of Ref. 96, but calculated cross section became less near factor 1.5-2 as it is seen on Fig. 12b (dashed curve).

Asymptotic constant for the second excited state 3/2⁻ of $^{13}$C calculated in Ref. 96 is 0.23 fm$^{-1/2}$, and recalculated dimensionless value is 0.24. It turned too difficult to find a potential able to reproduce AC value properly at 3.6845 MeV energy relatively GS of $^{13}$C nucleus or -1.26184 MeV relatively the threshold of n$^{12}$C channel. For getting the appropriate value of AC the potential might be very narrow, so the following found parameters describe AC approximately only

$$V_{3/2} = -681.80814 \text{ MeV}, \quad \gamma_{3/2} = 2.5 \text{ fm}^{-2}. \tag{34}$$

This potential gives the binding energy -1.261840 MeV with accuracy 10$^{-6}$ by FDM, charge radius 2.47 fm, mass radius 2.44 fm, and AC equals 0.30(1) within the interval 2-24 fm. It does not have FS, and reproduces the AC from Ref. 96 rather well.

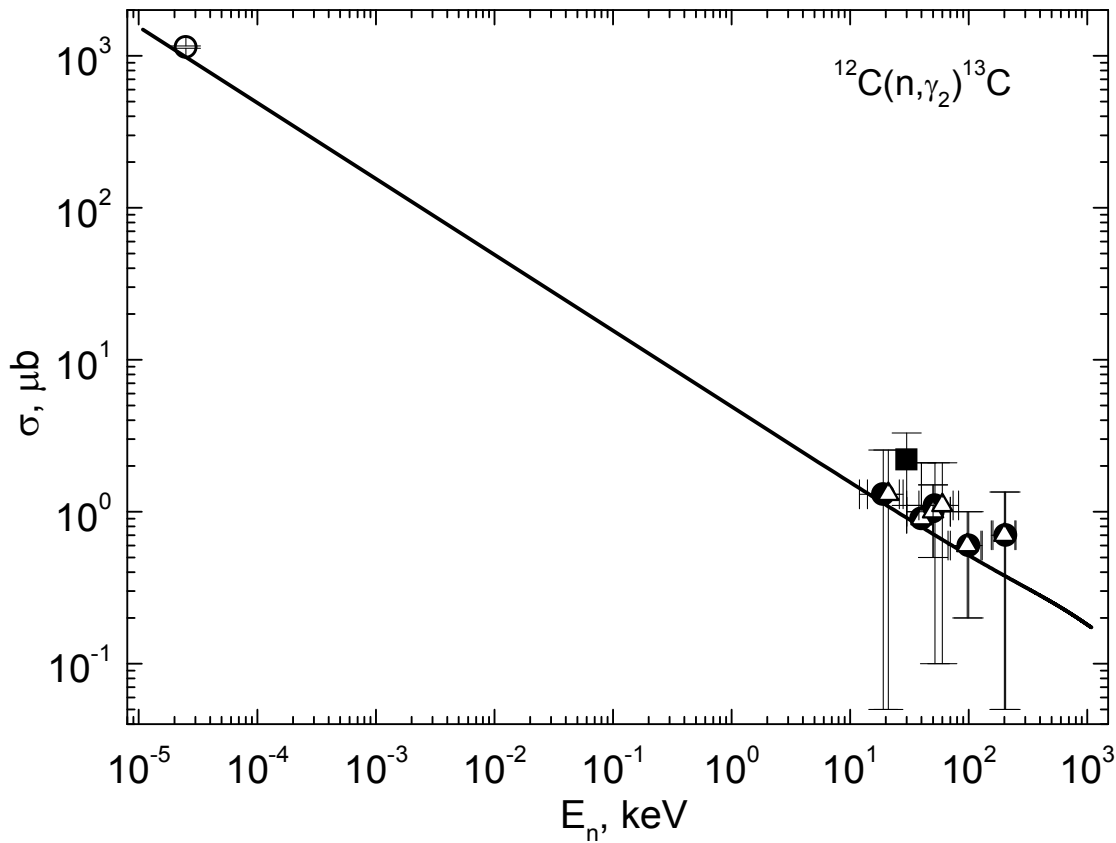

Fig. 12c. Total cross sections of the radiative neutron capture on $^{12}$C on the second excited state 3/2$^+$ of $^{13}$C at low energies. Experimental data: black squares (■) – Ref. 48, black dots (●) – Ref. 49, open triangles (Δ) – Ref. 47 and open circles (○) – Ref. 43. Solid line – total cross section calculated with ES potential of Eq (32) and scattering $P_{3/2}$ potential of Eq (34).

Comparison of experimental data and calculated cross sections of n$^{12}$C capture from the $^2S_{1/2}$ scattering state onto 3/2⁻ level done with potential of Eq. (32) are given on Fig. 12c. It is well seen that developing approach leads to good agreement in description of total cross sections in this case also. Thereby, the intercluster potentials here are coordinated with scattering phase shifts as usual and correctly describe the main characteristics of the considered BS of $^{13}$C in whole.

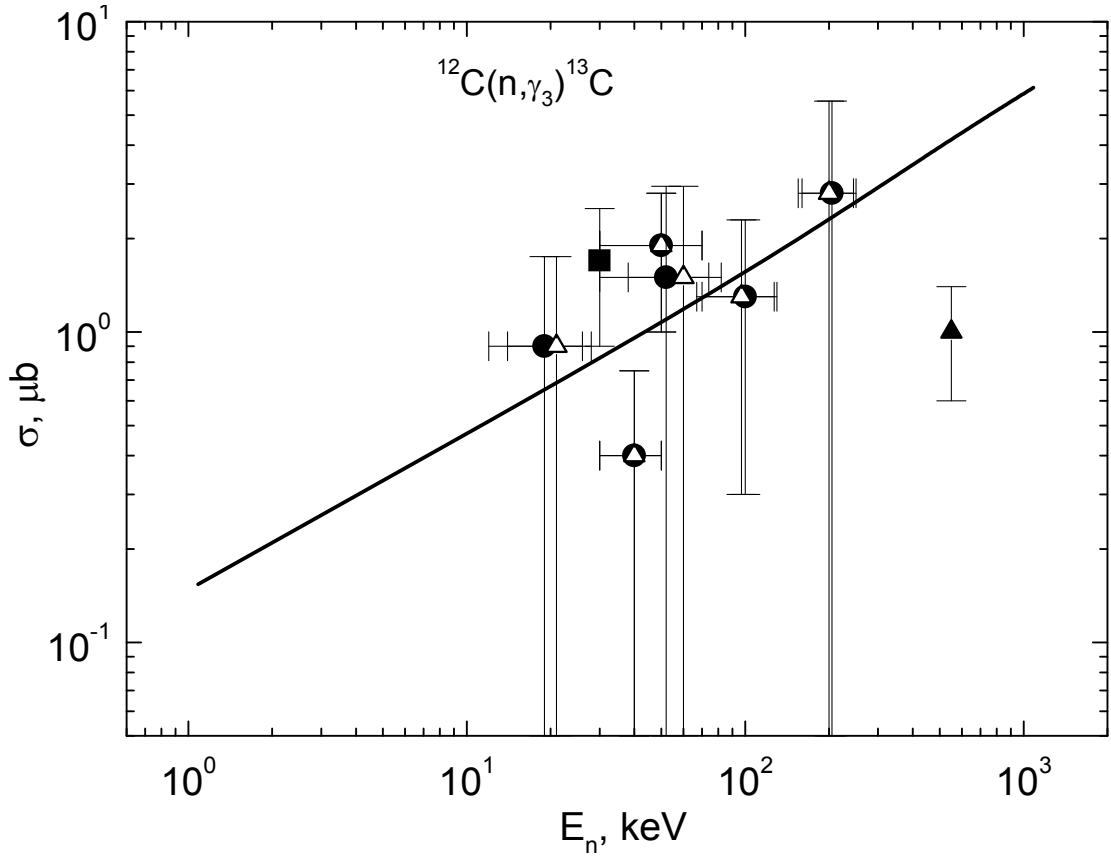

Fig. 12d. Total cross sections of the radiative neutron capture on $^{12}$C on the third excited state $5/2^+$ of $^{13}$C at low energies. Experimental data: black squares (■) – Ref. 48, black dots (●) – Ref. 49, open triangles (Δ) – Ref. 47, closed triangles ▲ – Ref. 45. Solid line – total cross section calculated with ES potential of Eq. (35) and scattering $P$ potential with zero depth.

For consideration of the transitions from $P_{3/2}$ scattering wave onto binding $D_{5/2}$ state at the energy -1.09254 MeV relatively the threshold of n$^{12}$C channel, the third excited state in $^{13}$C, let us present the AC. In Ref. 96 value 0.11 fm$^{-1/2}$ was obtained, and in Refs. 18-21 it is 0.15(1) fm$^{-1/2}$. Recalculated values at $\sqrt{2k} = 0.971$ are practically the same. For $P_{3/2}$ scattering wave zero potential was used as before. For the binding $D_{5/2}$ state a potential with one FS and same geometry as for the GS of $^{13}$C in Eq. (30) was used

$$V_D = -263.174386 \text{ MeV and } \gamma_D = 0.2 \text{ fm}^{-2}, \qquad (35)$$

It gives the binding energy -1.092540 MeV with accuracy 10$^{-6}$ by FDM, charge radius 2.49 fm, mass radius 2.61 fm, and AC equals 0.25(1) within the interval 6-25 fm. It has FS, and reproduces properly the order of magnitude of AC.

Comparison of experimental data and calculated cross sections of n$^{12}$C capture from the $^2P_{3/2}$ scattering state onto the binding $^2D_{5/2}$ level are given in Fig. 2d by the solid line together with experimental data. It is well seen that developing potential cluster model leads to quite reasonable agreement in description of total cross sections for the capture on the third ES of $^{13}$C also.

As at the energies from 10$^{-5}$ to 10 keV the calculated cross section $\sigma_{theor}$ is practically straight line (solid line in Fig. 12a) it may be approximated at low energies by simple function

$$\sigma_{ap}(\mu b) = \frac{12.7289}{\sqrt{E_n(keV)}}. \tag{36}$$

Constant value 12.7289 µb keV$^{1/2}$ is defined by one point of cross sections at the minimal energy 10$^{-5}$ keV. Modulus of relative deviation between the calculated $\sigma_{theor}$ and approximated $\sigma_{ap}$ cross sections

$$M(E) = \left|[\sigma_{ap}(E) - \sigma_{theor}(E)]/\sigma_{theor}(E)\right| \tag{37}$$

in the energy range from 10$^{-5}$ up to 10 keV is less than 1.0 %. We would like to assume the same energy dependence shape of the total cross section at lower energies. So, implemented estimation of cross section done at 1 µeV (10$^{-6}$ eV = 10$^{-9}$ keV) according Eq. (36) resulted 402.5 mb.

### 5.2. *Total cross sections for neutron capture on $^{13}C$*

Classification of the orbital states for the p$^{13}$C system, and hence the n$^{13}$C one by the Young schemes we did in Ref. 54. So, let us remind shortly that for p$^{13}$C system within the 1$p$ shell we got $\{1\} \times \{4441\} \rightarrow \{5441\} + \{4442\}$.[105] The first of the obtained scheme is compatible with the orbital momentum $L = 1$ only. It is forbidden as five nucleons can not occupy the $s$ shell. The second scheme is allowed and is compatible with the angular moments $L = 0$ and 2.[105] Thus, restricting by the lowest partial waves we conclude that there is no forbidden state in $^3S_1$ potential, but $^3P$ wave has both one forbidden and one allowed state. The last one appeared at the binding energy -8.1765 MeV of n$^{13}$C system[106] and corresponds the ground state of $^{14}$C nuclei in this channel with $J^\pi = 0^+$.

Note, as the isospin projection in n$^{13}$C system $T_z = -1$, then the total isospin $T=1$, and this is the first cluster system among all treated earlier ones pure by isospin with its maximum value.[3,54] Further, the $E$1 transition is taken into account at the consideration of the radiative n$^{13}$C $\rightarrow$ $^{14}$C$\gamma$ capture process, which is possible between the triplet $^3S_1$ scattering state onto the ground $^3P_0$ bound state of $^{14}$C in the n$^{13}$C channel. The nuclear part of the n$^{13}$C intercluster interaction is represented, as usual, in the Gaussian form without Coulomb part, for the calculation of total cross sections of the radiative capture.

As the $^3S_1$ wave potential without FS we used firstly the parameters fixed for the p$^{13}$C scattering channel[52,54]

$$V_S = -265.4 \text{ MeV}, \ \gamma_S = 3.0 \text{ fm}^{-2}, \tag{38}$$

Figure 13 shows the result of $^3S_1$ phase shift calculation (the dashed curve) with p$^{13}$C potential without Coulomb interaction, i.e., for the n$^{13}$C scattering system. It does not reveal now the resonance behavior,[52] but depends smoothly from energy. As there is no FS in this system this phase shift starts from zero value.

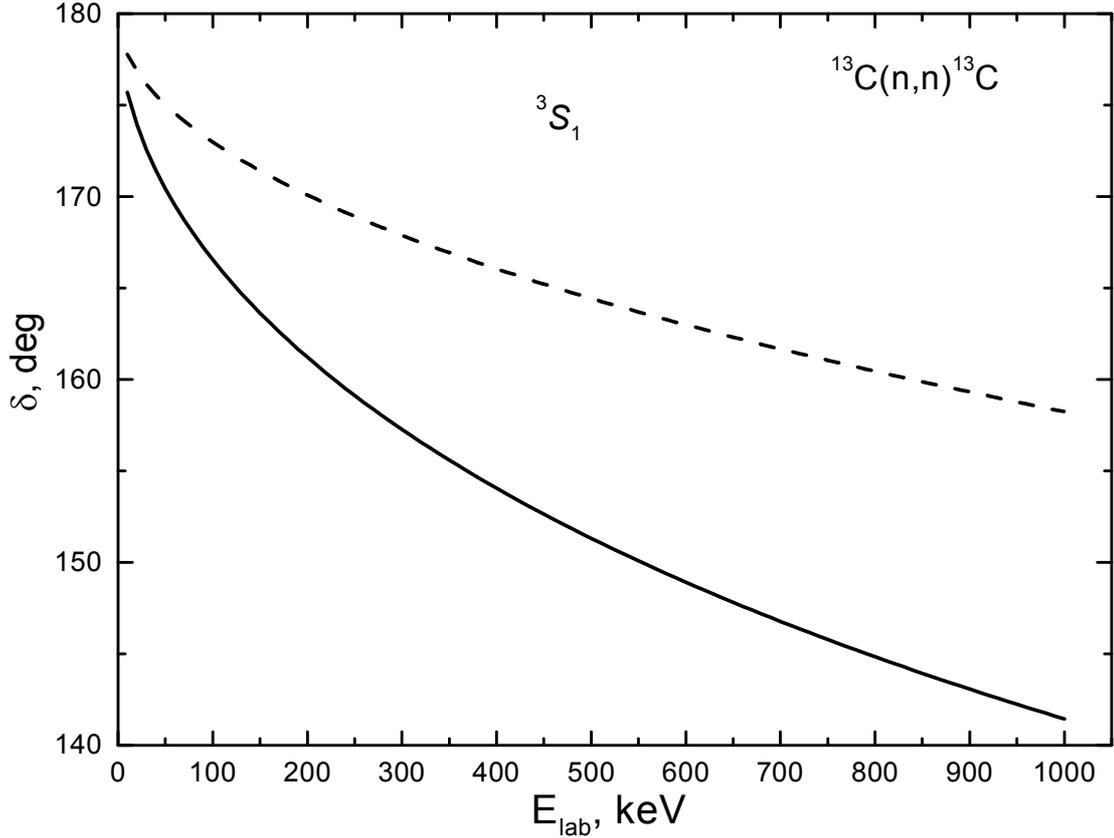

Fig. 13. Low energy $^3S_1$ phase shift of the n$^{13}$C elastic scattering. Dashed curve – calculations with potential of Eq. (38): solid – modified potential of Eq. (40).

Potential with one FS of triplet bound $^3P_0$-state should reproduce properly the binding energy of $^{14}$C in $J^\pi = 0^+$ ground state equals in n$^{13}$C-channel -8.1765 MeV Ref. 28 as well as describe the mean square radius of $^{14}$C according the experimental value 2.4962(19) fm.[106] The appropriate parameters have been obtained basing on the start set for $^{14}$N in bound p$^{13}$C-state

$$V_{\text{g.s.}} = -399.713125 \text{ MeV}, \quad \gamma_{\text{g.s.}} = 0.45 \text{ fm}^{-2}. \tag{39}$$

This potential gives the binding energy -8.176500 MeV with FDM accuracy $10^{-6}$ and charge mean square radius $R_{\text{ch}} = 2.47$ fm and mass radius 2.47 fm. For the asymptotic constant in dimensionless form[35] value 1.85(1) was obtained in the interval 4-12 fm being averaged by pointed above interval. Note, that for the AC in this channel value 1.81(26) fm$^{-1/2}$ was obtained in Ref. 111, and after the recalculation at $\sqrt{2k} = 1.02$ its dimensionless value 1.77(25) is in agreement with the present calculations.

As the additional computing control for the calculation of binding energy the variational method was applied.[61] It gave the energy value -8.176498 MeV with dimension $N = 10$ and independent parameter varying of the BS potential of Eq. (39). Varying parameters for the radial WF are given in Table 9, residuals do not exceed $10^{-11}$.[61] The charge radius and asymptotic constant do not differ from the above obtained in the FDM calculations. As it was said before, the average value -8.176499(1) MeV obtained by FDM and VM might be regarded as the true binding energy, i.e., the accuracy of determination of the binding energy of $^{14}$C in the n$^{13}$C channel by two methods and by different computer programs for the potential of Eq. (39) is on the level ±1.0 eV.

Pass to the description of our calculation results, let us note that experimental data on the total cross sections of the radiative neutron capture on $^{13}$C[43,49,50,112-114] were obtained by us from the Moscow State University data base[35] and Fig. 14 shows the pointed experimental data for the energies 25 meV÷100 keV.

Let us notice that the description of the total cross sections, as in work Ref. 54 too, we accounted only the $E1$ transition from the non-resonating scattering $^3S_1$ and $^3D_1$ waves obtained with central potential of Eq. (38) to the triplet $^3P_0$ bound state of $^{14}$C in the n$^{13}$C channel generated by the potential of Eq. (39). Calculated total cross sections for $^{13}$C(n, $\gamma_0$)$^{14}$C process at the energies lower 1.0 MeV with defined potential sets overestimate by near two orders the experimental data Refs. 49,50,112 in the region 10-100 keV.

Table 9. The variational parameters and expansion coefficients of the radial WF of $^{14}$C in n$^{13}$C channel for the GS potential of Eq. (39).

| $i$ | $\alpha_i$ | $C_i$ |
|---|---|---|
| 1 | 3.243302710972528E-002 | 5.549734166010744E-004 |
| 2 | 7.407629850544269E-002 | 8.788338009916163E-003 |
| 3 | 1.633437583034525E-001 | 5.468739372358444E-002 |
| 4 | 3.454184675560051E-001 | 2.115826076329647E-001 |
| 5 | 6.966337352505032E-001 | 6.564262690816171E-001 |
| 6 | 1.351318567304702 | 1.714603229965451 |
| 7 | 3.764749418123264 | -7.632588291509569 |
| 8 | 5.768787772876172 | -805611763353469 |
| 9 | 10.047525122607720 | -9.116142354803093E-002 |
| 10 | 38.232890649234430 | 5.255268476391280E-004 |

*Note.* The normalization of the function in the range 0÷30 fm equals $N = 0.9999999999999997$.

Experimental data Ref. 114 at 25 meV may be reproduced if take the potential depth

$$V_S = -215.77045 \text{ MeV}, \quad \gamma_S = 3.0 \text{ fm}^{-2}. \qquad (40)$$

for $^3S_1$-wave at the same geometry. The corresponding phase shift and total cross section are given by solid curves in Figs. 13 and 14, respectively. The scattering potential describes properly the bound $^3S_1$ level with $J^\pi = 1^-$ in n$^{13}$C channel at excited at 6.0938 MeV, and leads to the binding energy -2.08270 MeV relatively the threshold, charge and mass radii 2.47 fm, and AC equals 1.13(1) within the interval 2-22 fm. The situation here is similar to those in previous system when the subthreshold resonance $^3S_1$ state in p$^{13}$C system becomes bound one if the Coulomb interaction is switch off.

Consequently it is seen that slight change of potential depth coordinated with the energy of binding $^3S_1$ level allows to reproduce the experimental data on the total capture cross sections from 25 meV up to 100 keV (Fig. 14). Slowdown of the cross section at 0.5÷1.0 MeV is coming due to $E1$ transition from $^3D_1$ scattering wave which input is noticeable in this energy region only. Estimation of $M2$-transition from the resonating $^3P_2$ scattering wave corresponding to $J^\pi = 2^+$ at 141 keV in c.m. to $^3P_0$ ground state shows near 1% input from the $E1$ cross section.

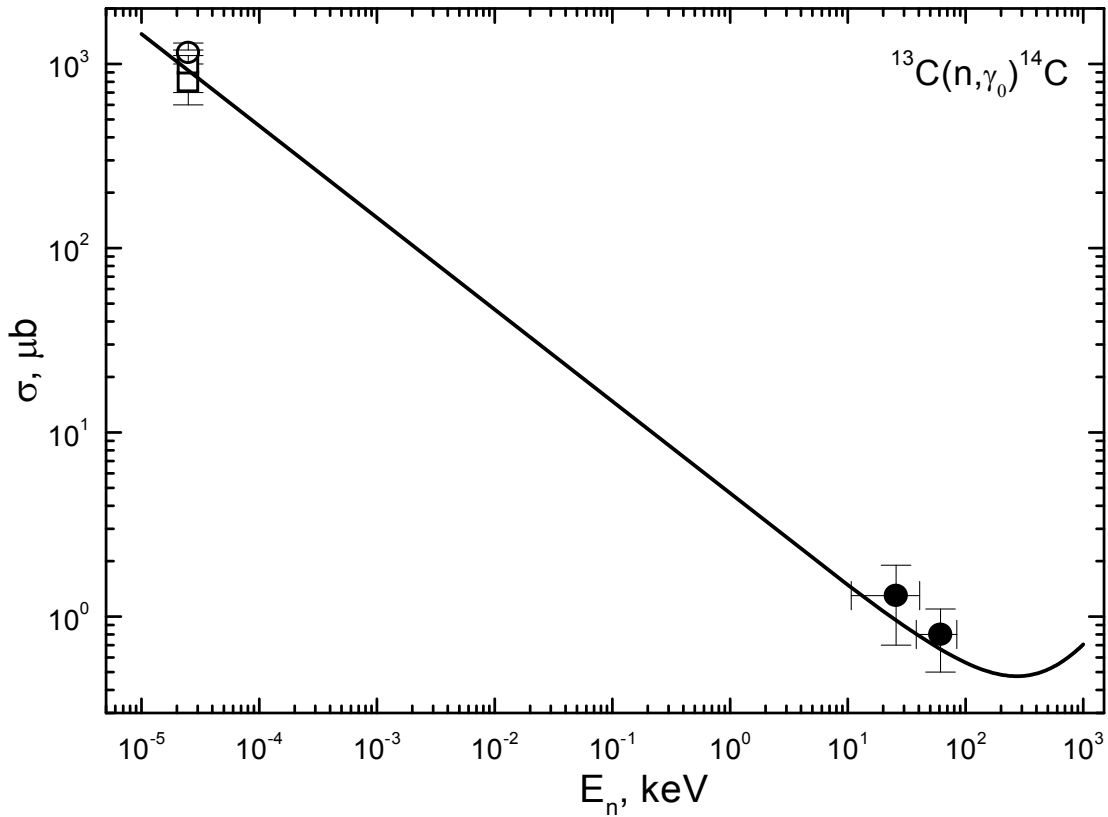

Fig. 14. – Total cross sections of the radiative neutron capture on $^{13}$C at low energies. Experimental data: ● – Refs. 114, 115, □ – Ref. 114, ○ – Ref. 43. Lines – calculations of total cross sections with potentials given in text.

Specially should be mentioned that comparing the previous n$^{12}$C system we did not find any independent information on AC for the first ES in binding $^3S_1$ wave. That is why scattering potential of Eq. (40) may have some ambiguity in parameters. We do not exclude that there might be another set of parameters which may describe correctly the characteristics of boundary state, in particular binding energy and total capture cross sections, but leading to somewhat another asymptotic constant.

We did not succeed in search of the n$^{13}$C phase shift analysis and experimental data on elastic scattering differential cross sections at the energies below 1.0 MeV. Available data above 1.26 MeV were measured with too large energy step[115] and do not allow to carry out the phase shift analysis, as it was done by us earlier, for example, for n$^{12}$C, p$^{12}$C and p$^{13}$C scattering at the energies, below 1.0 MeV.[52,53,108] Let us note that for the p$^{13}$C scattering in the resonance region at 0.55 MeV (l.s.) and its width 23(1) keV there have been near 30 measurements of differential cross sections done by several groups at four scattering angles. So detailed data allowed to reproduce within the phase shift analysis[52] the $J^\pi T = 1^-1$ resonance at 8.06 MeV relatively $^{14}$N ground state or 0.551(1) MeV relatively the threshold of the p$^{13}$C channel.[106]

In case of n$^{13}$C scattering data above 1.26 MeV on elastic differential cross sections[115] are given with too large energy step and are not appropriate for the reproducing even the shape of $J^\pi = 1^-$ resonance at 9.8 MeV relatively GS or at 1.75 MeV relatively the n$^{13}$C threshold. Parameters of this resonance are given in Table 14.7.[106] The situation with the resonance at 153 keV (l.s.) with width 3.4 keV (c.m.), which may be correspond to $J^\pi = 2^+$, is even worse. So, absence of reliable data on scattering differential cross sections at low energies leads to the 10÷15% ambiguity for the potential parameters appeared in calculations for the $^3S_1$ wave of the n$^{13}$C elastic scattering.

Thereby, the BS interaction of the n$^{13}$C system describes the main characteristics of the GS of $^{14}$C quite reasonable, so as it was obtained earlier for the p$^{13}$C channel of $^{14}$N.[54] But, the absence of results for phase shift analysis leads to the impossibility to do any certain and final conclusions concerning the depth of the elastic scattering potential in the $^3S_1$ wave. It seems that just its value influences to the calculation results for total cross sections of the radiative neutron capture on $^{13}$C, in the first place.

Therefore, the carrying out the new future measurements of the differential cross sections of the n$^{13}$C elastic scattering, from 0.1 MeV where the $2^+$ resonance at 153 keV is observed, and at least till the region of the $1^-$ resonance at 1.75 MeV, will take it possible to carry out new similar phase shift analysis. At that, the step of the measurements has to exceed 1/5÷1/7 of their width, as it was done, for example, in work Ref. 109 for the n$^{12}$C elastic scattering. In other case, it will be practically impossible to determine the form of resonance phase shifts of elastic scattering on the basis of phase shift analysis. The availability of these measurements, by-turn, can take the possibility more accurately obtain the characteristics of the corresponded partial potential of the elastic scattering and to carry out the single-valued calculations of the total cross sections of the radiative neutron capture on $^{13}$C.

Let us once more note in the concluding part of this section that, since at the energies from 10 meV to 10 keV the calculated cross section is shown in Fig. 14 by practically straight line, it may be approximated at low energies by simple function of the form of Eq. (7) with the constant value 4.6003 μb·keV$^{1/2}$. It is, as usual, defined by one point in the calculated cross sections at the minimal energy 10 meV. The absolute value $M(E)$ of Eq. (8) of relative deviation between the calculated cross section and its approximation of this function of Eq. (7) in the energy range from 25 meV to 10 keV is less than 0.4%. We would like to assume, as before, that the same energy dependence shape of the total cross section of Eq. (7) will be saved at lower energies too. The estimation of the cross section, for example, at 1 μeV (10$^{-6}$ eV = 10$^{-9}$ keV) gives the value 145.5 mb.

6. **Conclusion**

The calculations of the total cross sections of the radiative neutron capture on $^2$H at the energy from 10 meV to 15 MeV carried out in this paper and based on the principles of the PCM, in whole have a good agreement with the available experimental data. The potential cluster model, with the forbidden states and classifications of the orbital states of clusters according to the Young schemes that we used before, is able to correctly describe the general shape of the total cross sections of the neutron capture on $^2$H, at the same time with the description of the astrophysical $S$-factor of the proton capture on $^2$H.[3] Small changes of depth of the $^2S$ potential for this system are quite permissible, since the data of the p$^2$H phase shift analysis contain the big ambiguities and errors, and the data on the n$^2$H phase shift analysis, evidently, are absent at all. All of this leads to the certain ambiguity for parameters of the potentials of the n$^2$H interactions, which, as it was shown above, does not exceed 10%.[116]

The used potential cluster model and the intercluster potentials that are given above, as well as in the case of lighter nuclei,[3] allow to obtain a quite reasonable results during the description of the process of the radiative neutron capture on $^6$Li at the astrophysical energy range.[93] The results of the carried out calculations of the radiative neutron capture on $^6$Li, obtained only on the basis of the $E$1 transitions at the energy from 25 meV to 1.5 MeV in whole is in a good agreement with the available experimental data, as for the capture process, so as for the recalculated data for measurements of the total cross sections for two-particle photodisintegration of $^7$Li in the n$^6$Li channel.

For the n$^7$Li system, it is possible to describe the value of the total capture cross sections in the non-resonance energy range in the frame of the potential cluster model.[3] In addition, it is possible to describe the location and value of the $^5P_3$ resonance at low energy. It is possible to find parameters of the intercluster potentials for correct description of the total capture cross sections, which are corresponded with the given above classifications according to the orbital cluster states in this system. Consequently, the usage of the described conceptions about potentials with forbidden states, corresponded with the elastic scattering phase shifts of clusters and with the characteristics of the BS of $^8$Li, allows correctly describe the available experimental data for the radiative neutron capture on $^7$Li in the wide energy range.

Present results show that appropriate n$^{12}$C scattering potential of Eq. (32) coordinated with corresponding phase shifts[108] together with correct reproducing of $^{13}$C GS enable to describe the available experimental data on the radiative neutron capture cross sections at the energies from 25 meV to 100 keV. All potentials satisfied the classification of FS and AS by orbital Young schemes. Potential constructed for the GS reproduces the basic characteristics of $^{13}$C, i.e. binding energy in n$^{12}$C channel, mean square radius and asymptotic constant. So, these results may be regarded as one more confirmation of the success of cluster model approach applied early to the radiative neutron capture processes in other systems.[116] PCM succeeded also in description of radiative capture reactions of protons and other charge clusters on light nuclei.[2,4,5]

Constructed within PCM two-body n$^{13}$C potentials for $^3S_1$ wave and $^{14}$C GS, show good results for the total neutron radiative capture on $^{13}$C in the energy range from 25 meV up to 100 keV. Two-body potential used for the bound $n^{13}$C system reproduces well the basic GS characteristics of $^{14}$C, as well as it was done for $^{14}$N in p$^{13}$C channel.[54] At a time, it is rather difficult make a certain and final conclusions on potential depth for elastic scattering $^3S_1$ wave as there are essential ambiguity in available experimental radiative capture data. It seems this very value define the calculation results for the radiative neutron cross section capture on $^{13}$C. New measurements of differential cross sections for elastic n$^{13}$C scattering in the energy region up to 1.0 MeV with sufficient step might provide the careful phase shift analysis and define the shape of $^3S_1$ elastic phase. This may improve the definition of scattering potential and realize more unambiguous calculations of total radiative cross section capture on $^{13}$C.

Note, that this is already the twenty cluster systems, which was considered by us earlier on the basis of potential cluster model with the classification of the orbital states according to the Young schemes,[5] where it is possible to obtain the acceptable results for description of the characteristics of the radiative nucleon or light cluster capture processes on the $1p$ shell nuclei on the basis of the carried out classification of the cluster states. The properties of these cluster nuclei, their characteristics and considered cluster channels are listed in Table 7. The last results, shown in Table 10 and obtained in the frame of the PCM, are given here and in works Refs. 116-124.

Table 10. The characteristics of nuclei and cluster systems, and references to works in which they were considered.

| No. | Nucleus ($J^\pi, T$) | Cluster channel | $T_z$ | $T$ | Refs. |
|---|---|---|---|---|---|
| 1. | $^3$H ($1/2^+,1/2$) | $n^2$H | $-1/2 + 0 = -1/2$ | $1/2$ | 116 |
| 2. | $^3$He ($1/2^+,1/2$) | $p^2$H | $+1/2 + 0 = +1/2$ | $1/2$ | 3 |
| 3. | $^4$He ($0^+,0$) | $p^3$H | $+1/2 - 1/2 = 0$ | $0 + 1$ | 3 |
| 4. | $^6$Li ($1^+,0$) | $^2$H$^4$He | $0 + 0 = 0$ | $0$ | 11-13 |
| 5. | $^7$Li ($3/2^-,1/2$) | $^3$H$^4$He | $-1/2 + 0 = -1/2$ | $1/2$ | 11-13 |
| 6. | $^7$Be ($3/2^-,1/2$) | $^3$He$^4$He | $+1/2 + 0 = +1/2$ | $1/2$ | 11-13 |
| 7. | $^7$Be ($3/2^-,1/2$) | $p^6$Li | $+1/2 + 0 = +1/2$ | $1/2$ | 118 |
| 8. | $^7$Li ($3/2^-,1/2$) | $n^6$Li | $-1/2 + 0 = -1/2$ | $1/2$ | 85 |
| 9. | $^8$Be ($0^+,0$) | $p^7$Li | $+1/2 - 1/2 = 0$ | $0 + 1$ | 11-13 |
| 10. | $^8$Li ($2^+,1$) | $n^7$Li | $-1/2 - 1/2 = -1$ | $1$ | 117 |
| 11. | $^{10}$B ($3^+,0$) | $p^9$Be | $+1/2 - 1/2 = 0$ | $0 + 1$ | 11-13 |
| 12. | $^{10}$Be ($0^+,1$) | $n^9$Be | $-1/2 - 1/2 = -1$ | $1$ | 124 |
| 13. | $^{13}$N ($1/2^-,1/2$) | $p^{12}$C | $+1/2 + 0 = +1/2$ | $1/2$ | 51 |
| 14. | $^{13}$C ($1/2^-,1/2$) | $n^{12}$C | $-1/2 + 0 = -1/2$ | $1/2$ | 121 |
| 15. | $^{14}$N ($1^+,0$) | $p^{13}$C | $+1/2 - 1/2 = 0$ | $0 + 1$ | 54 |
| 16. | $^{14}$C ($0^+,1$) | $n^{13}$C | $-1/2 - 1/2 = -1$ | $1$ | 121 |
| 17. | $^{15}$C ($1/2^+,3/2$) | $n^{14}$C | $-1/2 - 1 = -3/2$ | $3/2$ | 122 |
| 18. | $^{15}$N ($1/2^-,1/2$) | $n^{14}$N | $-1/2 + 0 = -1/2$ | $1/2$ | 122 |
| 19. | $^{16}$N ($2^-,1$) | $n^{15}$N | $-1/2 - 1/2 = -1$ | $1$ | 123 |
| 20. | $^{16}$O ($0^+,0$) | $^4$He$^{12}$C | $0 + 0 = 0$ | $0$ | 119,120 |